\documentclass[aps,prd,showpacs,superscriptaddress,a4paper,nofootinbib]{revtex4}

\usepackage{latexsym}
\usepackage{amsmath}
\usepackage{amssymb}
\usepackage{ulem}
\usepackage{fancyhdr}
\usepackage{natbib}
\usepackage{url}
\usepackage{graphicx}
\usepackage{grffile}
\usepackage[vcentering,dvips]{geometry}
\usepackage{sidecap}

\begin{document}

\title{Observational cosmology using characteristic numerical
relativity: Characteristic formalism on null geodesics}

\author{P.~J. van der Walt}
\affiliation{ Department of Mathematics, Rhodes University,
Grahamstown 6140, South Africa }
\author{N.~T. Bishop}
\affiliation{ Department of Mathematics, Rhodes University,
Grahamstown 6140, South Africa }

\begin{abstract}
The characteristic formalism of numerical relativity is based on a
system of coordinates aligned with outgoing null cones. While these
coordinates were designed for studying gravitational waves,
they can also be easily adapted to model cosmological past
null cones (PNCs). Similar to observational coordinates in the
observational approach to cosmology, this then provides a model that
only makes use of information causally connected to an observer.
However, the diameter distance, which is used as a radial
coordinate, limits the model's cosmological application to the
region prior to the PNC refocussing. This is because after
refocussing, the diameter distance ceases to be a unique measure of
distance. This paper addresses the problem by introducing a metric
based on the Bondi-Sachs metric where the radial coordinate is
replaced by an affine parameter. A model is derived
from this metric and it is then shown how an existing numerical
scheme can be adapted for simulation of cosmological PNC behaviour.
Numerical calculations on this model are found to have the same
stability and convergence properties as the standard characteristic
formalism.
\end{abstract}


\maketitle

\section{Introduction}

The observational approach to cosmology is the endeavour to
reconstruct the geometry of the Universe using observations that do
not require the prior assumption of a cosmological model. Using this
approach, it was shown in \cite{ellis85} that given ideal
cosmological observations, the only essential assumption necessary
to determine the geometry of the Universe is a theory of gravity.
Assuming General Relativity, then, the full set of Einstein field
equations (EFEs) can be used to reconstruct the geometry of the
Universe using direct observations on the past null cone (PNC) as
boundary conditions. Observationally and theoretically this is a
very ambitious task and therefore, current developments have been
restricted to spherically symmetric dust models while only relaxing
the usual assumption of homogeneity in the radial direction. These
restricted models are important for the development of theoretical
foundations and also useful as verification models since they avoid
the circularity of verifying what has already been assumed. For
instance, doing investigations such as quantifying homogeneity on
different scales, testing the verifiability of cosmology
\cite{ellis85}, validating the Copernican principle \cite{uzan08}
and determining the metric of the Universe \cite{hellaby09} require
more general models than the conventional Robertson-Walker
geometries.

The \emph{Observational Cosmology} (OC) programme of Ellis and
others played an instrumental part in the development of the
observational approach. Of particular interest is their use of
observer coordinates which is based on the optical coordinates of
Temple \cite{temple38}. These coordinates are based on the natural
propagation of electromagnetic radiation and follow the causal
structure of an observer PNC. This approach then consists of two
problems: firstly, astronomical observations are used to set up the
metric on the local PNC and secondly these form the final values of
a characteristic final value problem which determines the historical
evolution of the region causally connected to the PNC (i.e. the
interior of the PNC). The PNC and its interior, which is the
causally connected region of a cosmological observer, is of
fundamental importance since it defines the limits on which
cosmological models can be validated from direct observations. Apart
from the initial \emph{Ideal Observational Cosmology} report
\cite{ellis85}, which focussed on the fundamental principles, the
focus of the OC programme has mostly been on exact solutions and
perturbations using tetrad formalisms in spherical symmetry. Some
recent developments are presented in
\cite{araujo00,araujo09,araujo10} and a summary in \cite{araujo11}.
Numerical algorithms have been proposed in \cite{hellaby09} using a
metric based approach but these have not been implemented. A
separate line of development based on the Lema\^{i}tre-Tolman-Bondi
(LTB) model, using comoving coordinates, using the models introduced
in \cite{musta98.1,musta98.2}, was followed in \cite{lu07,mcclure08}
to reconstruct the observer metric numerically. Some important
aspects addressed in the LTB approach are the treatment of a
reconverging PNC through coordinate singularities and the
sensitivity of the numerics when using realistic data as boundary
conditions.

The characteristic formalism of numerical relativity~\cite{bish97,
Winicour09,Reisswig:2009us} usually associated with studying
gravitational waves, was put forward in \cite{bishop96,
vanderwalt10} as a method for modelling the observable universe.
With the motivation that there exists a well-established base of
null-cone numerical schemes, it was investigated to what extent
these can be employed to simulate the evolution into the past of a
cosmological PNC, instead of modelling the evolution into the future
of a future null cone. It was found that with minor cosmological
considerations, this method was very well suited for modelling the
historical evolution of the observable universe. That is apart from
a known limitation of the characteristic formalism; i.e. in
modelling the reconvergence of the PNC of an expanding universe, the
diameter distance, as radial coordinate, becomes multi-valued and
limits the feasible radial extent of this formalism to the region
prior to the observer apparent horizon (AH). Since the properties of
the observable universe at the AH are directly related to the
cosmological constant
($\Lambda$)~\cite{hellaby06,araujo09,hellaby09}, extending the model
beyond the AH is a compulsory requirement for studying contemporary
problems in cosmology.

This paper addresses this problem by introducing a metric based on
the Bondi-Sachs metric where the radial coordinate is replaced with
an affine parameter. We derive the model with a cosmological constant
$\Lambda$ incorporated into the Einstein field equations, and regard
$\Lambda$ as a parameter of the theory of gravity rather than as a
matter source term. Similar to
the conventional characteristic formalism, this model consists of a
system of differential equations for numerically evolving
the EFEs as a characteristic initial value problem (CIVP). A
numerical code implemented for the method has been found to be
second order convergent. This code enables simulations of different
models given identical data on the initial null cone and provides a
method to investigate their physical consistency within the causally
connected region of our current PNC. These developments closely
follow existing 3D schemes developed for gravitational wave
simulations, which should make it natural to extend the affine CIVP
beyond spherical symmetric simulations.

In section \ref{sec:affine.model}, an affine parameter is introduced
in the Bondi-Sachs metric to derive the affine CIVP model for
cosmology. Details of the numerical implementation are given in
section \ref{sec:affine.nummerical} and coordinate transformations,
which relate comoving coordinates to the affine CIVP coordinates are
discussed in section \ref{sec:affine.transform}. Results of
simulations with the new coordinates are presented in section
\ref{sec:affine.results}. Section \ref{sec:affine.conclusion}
concludes the paper.


\section{Characteristic formalism on null geodesics}
\label{sec:affine.model}

\subsection{Characteristic formalism with affine radial coordinate}
\label{sec:affine.mod}

The conventional characteristic formalism in numerical relativity
uses a frame of reference based on outgoing null cones that evolve
from values on an initial null cone. The idea is conceptualised in
figure \ref{sec:obs-coords}. $G$ is a timelike geodesic, and $u$ is
the proper time on $G$. (In practice, in numerical relativity simulations
the inner boundary of the null cone is usually a timelike worldtube
rather than a geodesic, but the problem can be formulted with a
geodesic). Null geodesics emanating from $G$ have
constant $(u,\theta,\varphi)$, and near $G$ the angular coordinates
$\theta$ and $\varphi$ have the same meaning as in spherical polar
coordinates. The coordinate $r$ is the diameter, or area, distance
defined by the condition that the surface area of a shell of
constant $r$ is $4\pi r^2$.

The geometry of the characteristic formalism is described by the
Bondi-Sachs metric, which in spherical symmetry is formulated as
\footnote{The notation used here is based on that of \cite{bish97}
and substituting $W=V-r$ will give the original notation of Bondi
and Sachs in \cite{bondi60,bondi62,sachs62}.}
\begin{align}
ds^2 = -e^{2\beta} \left(1+\frac{W}{r} \right) du^2
        - 2 e^{2\beta} dudr + r^2 \{ d \theta ^2
        +  \sin ^2 \theta d \varphi ^2\} . \label{sec:civp-met}
\end{align}
This can be recognised as a generalization of the well-known
Eddington-Finkelstein form of the exterior Schwarzschild metric,
obtained by setting $\beta=0, W=-2M$ where $M$ is the mass of the
source. The coordinate system is defined such that $\beta$ and $W$
vanish at the vertex of each null cone, i.e. at $r=0$ equation
(\ref{sec:civp-met}) reduces to a Minkowskian metric.

Starting with the Bondi-Sachs metric (\ref{sec:civp-met}), an
affinely paramterized geodesic in these coordinates is determined
through
\begin{align}
\frac{d^2 r}{d \lambda ^2} + \Gamma_{11} ^{1}
\left(\frac{dr}{d\lambda}\right)^2 &= 0 \Rightarrow \frac{d^2 r}{d
\lambda ^2} + 2 \beta_{,r} \left(\frac{dr}{d\lambda}\right)^2 = 0 .
\end{align}
Setting $\lambda=r$ at the origin provides the initial conditions
$r(0) = \lambda(0) = 0$ and $dr/d\lambda|_{\lambda=0}=1$ then
solving gives
\begin{align}
\frac{dr}{d\lambda} = e^{-2 \beta} . \label{sec:affine-drdl2}
\end{align}

Using tensor transformation laws and substituting
(\ref{sec:affine-drdl2}) for all $\partial r/\partial \lambda$
terms, a new metric with the radial coordinate $\lambda$ is
introduced as
\begin{align}
ds^2 &= -\left(1+\frac{\hat{W}}{\hat{r}}\right) du^2
    - 2 dud\lambda + \hat{r}^2 \{d\theta ^2
    + \sin ^2 \theta d \varphi ^2\} \label{sec:affine.met} \\
\textrm{with:~~} \hat{W} &= \hat{W}(u,\lambda) \textrm{~~and~~}
    \hat{r} = \hat{r}(u,\lambda) . \nonumber
\end{align}
Substituting (\ref{sec:affine.met}) into the EFEs, using the form
$R_{ab} = \kappa(T_{ab} - \frac{1}{2} T g_{ab}) + \Lambda g_{ab}$,
with the stress-tensor for a dust-like fluid ($T_{ab} = \rho v_a
v_b$ and $T=-\rho$) gives \footnote{Using geometric units $G=1, c=1$
and $\kappa=8 \pi$}
\begin{align}
\hat{r}_{,\lambda \lambda} &=
    -\frac{1}{2} \kappa \hat{r} \rho (v_{1})^{2} \label{sec:affine.rll} \\
\hat{r}_{,u \lambda} &=
    \frac{1}{2}
    \left\{\hat{W}_{,\lambda} \hat{r}_{,\lambda}
    + \hat{r}\hat{r}_{,\lambda \lambda}
    + \hat{W} \hat{r}_{,\lambda \lambda}
    - 2 \hat{r}_{,u} \hat{r}_{,\lambda}
    - 1
    + (\hat{r}_{,\lambda})^2
    + \frac{1}{2} \kappa \rho \hat{r}^2
    + \Lambda \hat{r}^2 \right\}
    \Big{/} \hat{r} \label{sec:affine.rul} \\
\hat{W}_{,\lambda \lambda} &=
    \frac{\hat{W}}{\hat{r}} \hat{r}_{,\lambda \lambda}
    + 4 \hat{r}_{,u \lambda}
    + 2 \kappa \left(v_0 v_1 \rho - \frac{1}{2} \rho \right) \hat{r}
    - 2 \Lambda \hat{r} \label{sec:affine.Wll} \\
\textrm{with:~~} \hat{r}(0) &=
    \hat{W}(0) =
    \hat{W_{,\lambda}}(0) =
    \hat{r}_{,u}(0) = 0
    \textrm{~~and~~}
    \hat{r}_{,\lambda}(0) = 1. \nonumber
\end{align}

Further, substituting the dust stress-tensor and
(\ref{sec:affine.met}) into the continuity equation,
$T^{ab}_{\;\;;b}=0$, the energy-momentum equations follow
\begin{align}
v_{1,u}  & = \frac{1}{v_{1}} \left\{
    \left(\hat{V}_w v_1 - v_{0}\right) v_{1,\lambda}
    + \frac{1}{2} (v_{1})^2 \hat{V}_{w,\lambda}
 \right\} \label{sec:affine.v1u} \\
\rho_{,u}  & = \frac {1}{v_1} \left\{ \rho \left[
    \hat{V}_w\left(  \frac{2v_{1}}{\hat{r}} \hat{r}_{,\lambda}
    + v_{1,\lambda}\right)
    - \left( \frac{2 v_{0}}{\hat{r}} \hat{r}_{,\lambda}
    + v_{0,\lambda} \right)
    + \hat{V}_{w,\lambda} v_{1}
    - \left(\frac{2 \hat{r}_{,u}}{\hat{r}}\right) v_{1} \right] \right. \nonumber \\
    & \hspace{45pt} \left.
    + \; \rho_{,\lambda} \left(\hat{V}_w v_{1} - v_0
    \right)
    - \rho v_{1,u} \right. \Bigg{\}} \label{sec:affine.rhou} \\
\textrm{with:~~} \hat{V}_w & = 1+ \frac{\hat{W}}{\hat{r}} .
\nonumber
\end{align}
Making use of the normalisation condition, $g^{ab}v_{a}v_{b}=-1$,
$v_{0}$ can be written in terms of $v_{1}$ as
\begin{align}
v_0 = \frac{1}{2} \hat{V}_w v_1 + \frac{1}{2} v_1^{-1}
\label{sec:affine.v0} .
\end{align}

Having the values on the initial null cone for $\rho$ and $v_1$,
equations (\ref{sec:affine.rll}) to (\ref{sec:affine.v0})
form a hierarchical system that can be solved in the order
(\ref{sec:affine.rll}), (\ref{sec:affine.rul}),
(\ref{sec:affine.Wll}) and (\ref{sec:affine.v0}), then solving
equations (\ref{sec:affine.v1u}) and (\ref{sec:affine.rhou}) evolves
the system to the next null cone where the process can be repeated
until the domain of calculation has been covered. These equations
are all interdependent and require an iterative scheme for a
numerical solution.

\subsection{Cosmological considerations}

\begin{figure}[h!]
\begin{center}$
\begin{array}{ll}
\hspace{0pt}
\includegraphics[width=0.4\textwidth]{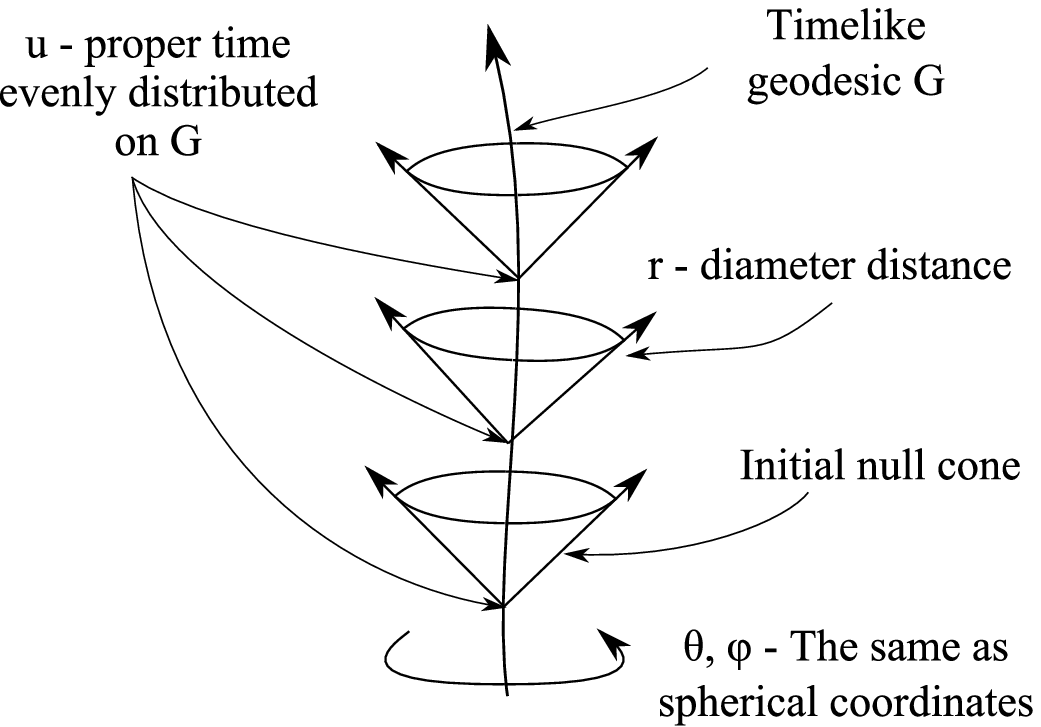} &
\hspace{10pt}\includegraphics[width=0.35\textwidth]{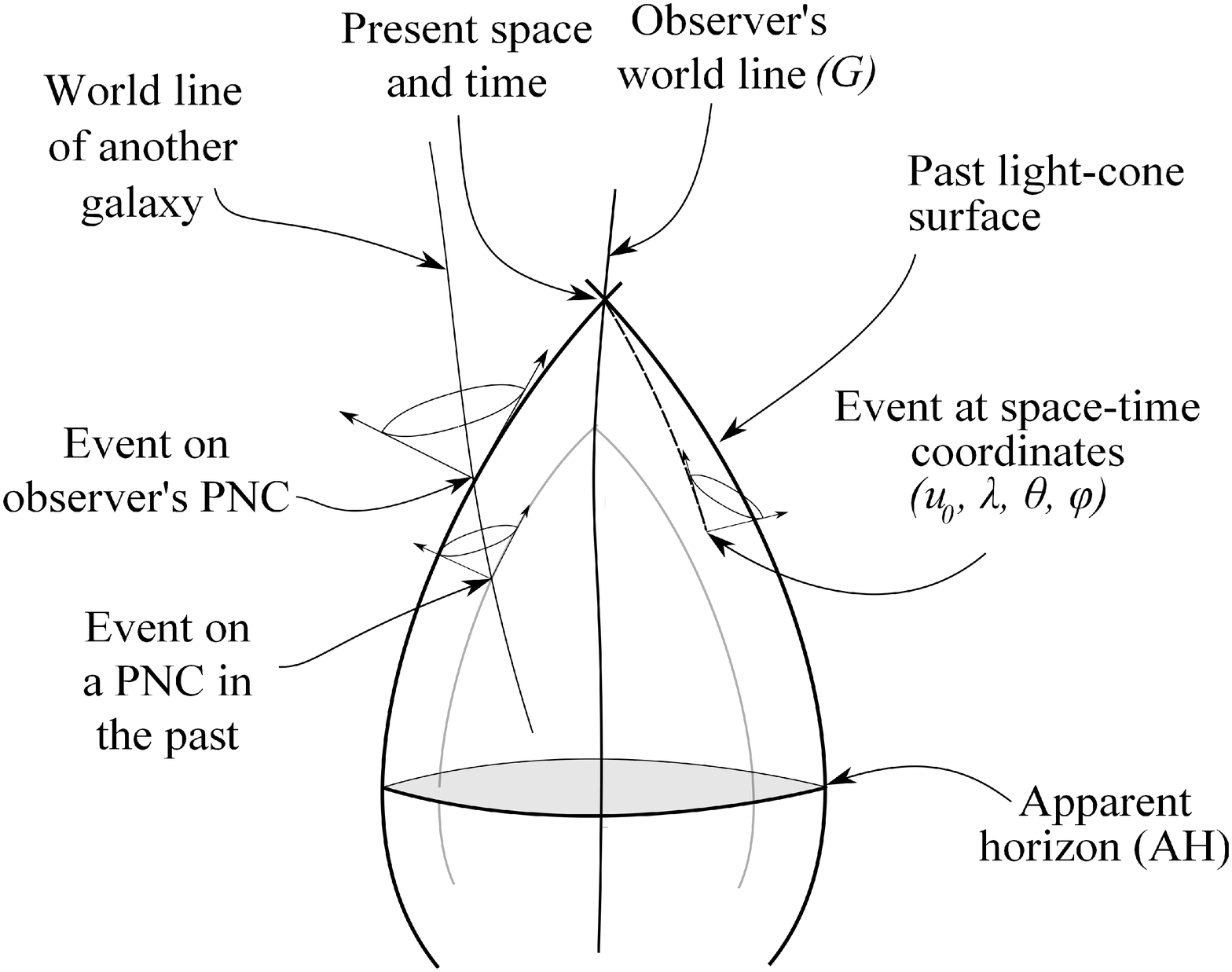}
\end{array}$
\end{center} \vspace{0pt} \caption{On the left hand side, the null
coordinates of the characteristic formalism and on the right, the
affine null cone coordinates for cosmology. } \label{sec:obs-coords}
\end{figure}

The model introduced in section \ref{sec:affine.mod} is essentially
a null cone formalism which makes provision for a null cone that can
reconverge at some distance from the cone vertex. Taking $G$ to be
the worldline of an observer located at the cone vertex and
integrating into the past, these coordinates can be naturally
aligned with a cosmological PNC. Further, it being spherically
symmetric and radially inhomogeneous, classifies it as a LTB model
in null coordinates. Using this model, the simulations done in
\cite{vanderwalt10} and \cite{bishop96} can be extended beyond the
apparent horizon (AH). Figure \ref{sec:obs-coords} illustrates the
differences between conventional characteristic coordinates and the
affine coordinates. Besides being not comoving, the affine
coordinates closely resemble the observer coordinates as described
in \cite{ellis85}.

As an observational cosmology problem, where the geometry is
determined from direct observations, there are two subproblems to be
solved:
\begin{itemize}
    \item[i.] Reconstruct the geometry of the local PNC from directly
    observable quantities, such as redshift-distance and galaxy number counts.
    \item[ii.] Evolve the model as a reverse CIVP to determine the interior of the
    PNC using the values determined in (i) as initial conditions.
\end{itemize}

Although the accumulation of cosmological data in recent years has
been astounding, these are not yet sufficiently complete for a
practical implementation of the first problem, which will have to
incorporate aspects such as data reduction and the sensitivity of
the model to observational errors. These aspects will not be
considered in this paper and only a conceptual description of the
relations between observable quantities and the required input for
the model will be described in the next section. The second
subproblem will be treated in detail in the remainder of the paper
through the development of a numerical code for the affine CIVP
model. By itself this provides a mechanism for testing hypothesised
models by evaluating the behaviour of their PNCs under given initial
values. The PNC behaviours of different models given similar initial
values are particularly interesting cases to investigate.

\subsection{Reconstructing the metric}
\label{sec-obs}

As initial data for the model, $v_1(\lambda)$ and $\rho(\lambda)$
have to be determined from observations on the PNC. These
require measures of the radial distribution of expansion and density.
As a measure
of expansion, redshift in terms of the luminosity distance ($d_L$)
can be determined from redshift-magnitude observations (e.g. type Ia
supernovae observations). The reciprocity theorem can then be used
to convert $d_L$-redshift to diameter distance-redshift relations
(see \cite{ellis71})
\begin{align}
z(d_L) \textrm{~~and~~} d_L = (1+z)^2 r_0 \Rightarrow \hat{z} =
z(r_0).
\end{align}
Here the zero subscript refers to a value on the current PNC.
The redshift is directly related to the time
component of the contravariant velocity:
\begin{align}
1+\hat{z} = \frac{d u}{d \tau} = v^0
\end{align}
where $\tau$ is the proper time along general galactic world lines
(see \cite{ellis85}). This can be used to determine the covariant
velocity $v_1$:
\begin{align}
v_1(r_0) = - v^0 = - (1+\hat{z}) . \label{sec:civp-v10}
\end{align}

As a measure of density, observed galaxy number counts ($n$) in
terms of $z$ can be used to determine $\rho(r_0)$. Galaxy redshift
surveys can provide data for this, although current surveys are not
yet sufficiently complete at high redshifts. Since
$\hat{z}(r_0)$ is known, $n(z)$ can be rewritten as $\hat{n}(r_0) =
n(\hat{z}(r_0))$. It was shown in \cite{bishop96} that the proper
number count can be written as
\begin{align}
N = \frac{n}{(1+z)} .
\end{align}
In terms of the diameter distance, the proper number count then
becomes
\begin{align}
\hat{N}(r_0) = \frac{\hat{n}}{(1+\hat{z})} .
\end{align}
The proper density is then related to the proper number count
\begin{align}
\rho(r_0) = f(\hat{N}). \label{sec:civp-rho0}
\end{align}
The details of this relation will not be considered at this stage
but in principle it must take into account aspects such as dark
matter and source evolution, preferably with factors independent of
an already assumed cosmological model.

From (\ref{sec:civp-v10}) and (\ref{sec:civp-rho0}) the required
values of $\rho$ and $v_1$ are obtainable in terms of the diameter
distance. Since these are required in terms of $\lambda$,
interchanging the independent variable in (\ref{sec:affine.rll})
provides a method to relate $\lambda$ in terms of $r$. Starting with
(\ref{sec:affine.rll}), rewritten as a system of ordinary
differential equations (ODEs), introducing $\hat{S}$,
\begin{align}
\hat{r}_{,\lambda} &= \hat{S} \\
\hat{S}_{,\lambda} &=
    \hat{r}_{,\lambda \lambda} =
    -\frac{1}{2} \kappa \hat{r} \rho (v_{1})^{2} \label{sec:obs-dSdl} \\
\textrm{with:~~} \hat{r}(0) &= 0 \textrm{~~and~~}
    \hat{S}(0) = 1, \nonumber
\end{align}
then interchanging the roles of $\lambda$ and $r$, simplifying,
rearranging and introducing $\hat{U}$ provides ODE expressions for
$\lambda_{, \hat{r}\hat{r}}$
\begin{align}
\lambda_{, \hat{r}} &= \hat{U} \\
\hat{U}_{, \hat{r}} &= \lambda_{, \hat{r}\hat{r}} =
    \left(\frac{1}{2} \kappa \hat{r} \rho (v_{1})^{2} \right)\hat{U}^3 \label{sec:obs-dUdr} \\
\textrm{with:~~} \lambda(0) &= 0 \textrm{~~and~~}
    \hat{U}(0) = 1. \nonumber
\end{align}

Although these equations will not be solved at this stage, there are
some complications that will require special consideration around
the AH. These arise from the fact that the diameter distance in an
expanding universe is not necessarily monotonically increasing and
therefore, not a unique independent variable. In these cases,
$\hat{r}$ reaches a maximum and then decreases. At the maximum
diameter distance, the AH, $\hat{r}_{,\lambda}=0$, and $\lambda_{,
\hat{r}}$ is singular. In general, this can be overcome by
separating the solution into regions prior, around and succeeding
the AH where the region around the AH is solved using series
expansions. Such a method was previously implemented in \cite{lu07}
to handle singularities around the AH for LTB models.

It should be noted that~\cite{hellaby06}, followed by \cite{araujo09}
and \cite{hellaby09}, derived a relationship for the cosmological constant
$\Lambda$ (which we are regarding as a parameter of the theory of gravity)
involving the maximum value of $\hat{r}$ and matter data within the PNC.
In order to test the theory of gravity more generally, a measurement involving a
rate of change over time is required. Redshift drift $z_{,u}(z)$ \cite{uzan08}, which is one of
the design objectives of the CODEX spectrograph planned for the
European Extremely Large Telescope (E-ELT) \cite{Pasquini10}, can be
used for this purpose. From (\ref{sec:civp-v10}) we have
\begin{align}
v_1(r_0) = - (1+\hat{z}) \Rightarrow v_{1,u} (r_0) =
-\hat{z}_{,u}(\hat{z}),
\end{align}
and then an evaluation of the difference between the left and right hand sides of
equation (\ref{sec:affine.v1u}) provides a test of the theory being used.

In the LTB models investigated in \cite{lu07} and \cite{mcclure08},
data reduction methods were introduced which provide valuable
insight into the methodology of converting realistic data into
useful initial data. In combination with the methods introduced in
this section, all options will have to be considered when data of
sufficient completeness becomes available for observational models.

\section{Numerical implementation}
\label{sec:affine.nummerical}

The numerical scheme described here is in structure similar to the
finite difference scheme described in \cite{vanderwalt10}, which in
turn was based on the general 3D code developed in in \cite{bish97}
and \cite{bish99}. Since the original scheme is based on first order
ODEs, the second order hypersurface equations are rewritten by
introducing $\hat{R}(u, \lambda) = \hat{r}_{,u}$, $\hat{S}(u,
\lambda)$ and $\hat{T}(u, \lambda)$, which then gives:
\begin{align}
\hat{r}_{,\lambda} &= \hat{S} \label{sec:affine.rl}\\
\hat{S}_{,\lambda} &=
    \hat{r}_{,\lambda \lambda} =
    -\frac{1}{2} \kappa \hat{r} \rho (v_{1})^{2} \label{sec:affine.Sl} \\
\hat{R}_{,\lambda} &=
    \hat{r}_{,u \lambda} =
    \frac{1}{2}
    \left\{\hat{T} \hat{S}
    + \hat{r}\hat{S}_{,\lambda}
    + \hat{W} \hat{S}_{,\lambda}
    - 2 \hat{R} \hat{S}
    - 1
    + \hat{S}^2
    + \frac{1}{2} \kappa \rho \hat{r}^2
    + \Lambda \hat{r}^2 \right\}
    \Big{/} \hat{r} \label{sec:affine.Rl} \\
\hat{W}_{,\lambda} &= \hat{T} \label{sec:affine.Wl}\\
\hat{T}_{,\lambda} &=
    \hat{W}_{,\lambda \lambda} =
    \frac{\hat{W}}{\hat{r}} \hat{S}_{,\lambda}
    + 4 \, \hat{R}_{,\lambda}
    + 2 \kappa \left(v_0 v_1 \rho - \frac{1}{2} \rho \right) \hat{r}
    - 2 \Lambda \hat{r}  \label{sec:affine.Tl} \\
\textrm{with:~~} \hat{r}(0) &=
    \hat{W}(0) =
    \hat{R}(0) =
    \hat{T}(0) = 0
    \textrm{~~and~~}
    \hat{S}(0) =1. \nonumber
\end{align}
These equations can be solved using standard ODE methods for systems
of equations where (\ref{sec:affine.rl}) and (\ref{sec:affine.Sl})
are solved as one system and (\ref{sec:affine.Rl}),
(\ref{sec:affine.Wl}) and (\ref{sec:affine.Tl}) are solved as a
separate system. In both cases a number of iterations are required
to obtain convergence.

The discretisation strategy followed is based on a rectangular grid
similar to the one used in \cite{vanderwalt10} but using $\lambda$
as the radial coordinate. Solving the hypersurface equations is done
with a central difference method on half steps between the $r$-grid
points, using
\begin{align}
g_{j}^i & = g_{j-1}^i + \frac{\Delta \lambda}{2}(g_{,r \;j}^i +
g_{,r\;j-1}^i)
\end{align}
with $i$ being the time step and $j$ the radial step. Here
$g_{,r\;j}^i$ is calculated by substituting known values into
equations (\ref{sec:affine.rl}-\ref{sec:affine.Tl}). In order to
solve the evolution equations, (\ref{sec:affine.v1u}) and
(\ref{sec:affine.rhou}), their general form is notated as
\begin{align}
v_{1,u} = F_{v1} \textrm{~~and~~} \rho_{,u} = F_{\rho}
\label{sec:code-gu}
\end{align}
and as explicit finite differences on a time half step they are
written as
\begin{align}
v_{1j}  ^{n+1} = v_{1j} ^{n} + \Delta u F_{v1\;j} ^{n+1/2}
\textrm{~~and~~} \rho_{j}  ^{n+1} = \rho_{j} ^{n} + \Delta u F_{\rho
j} ^{n+1/2} \label{sec:code.gun}.
\end{align}
Here, $n$ is a time iterator that will approach $i$. In these
equations, the numerical values at the point $(i,j)$ are used to
evaluate the matter terms and hypersurface derivatives. Radial
matter derivatives are calculated making use of standard central
difference formulae (see for instance \cite{burden93} p.160-161).

After setting up a suitable grid, the numerical algorithm can be
summarised in the following steps:
\begin{itemize}
\item[i.] Set the $\rho$ and $v_1$ initial values on to the initial grid
points. These values will, in principle, be obtained from
observations.
\item[ii.] Calculate $r$, $r_{,u}$, $W$ and $v_0$ from $\rho$ and $v_1$ on the initial null cone
using (\ref{sec:affine.rl}-\ref{sec:affine.Tl}).
\item[iii.] Calculate $F_{j} ^n$, with $n=1$ for the initial step,  from the values of $v_1$, $\rho$
, $r$, $r_{,u}$, $W$ and $v_0$ using (\ref{sec:code-gu}).
\item[iv.] Set $F_{j} ^{n+1/2} = F_{j} ^n$, again $n=1$ for the initial step, and calculate $v_1$ and
$\rho$ as an initial approximation that will approach the actual
values with subsequent iterations using (\ref{sec:code.gun}).
\item[v.] Use the new values of $v_1$ and $\rho$ to calculate
$r$, $r_{,u}$, $W$ and $v_0$ and their radial derivatives similar to
(ii).
\item[vi.] Calculate $F_{j}^{n+1/2} = 1/2\,(F_{j}^{n}+F_{j}^{n+1})$ from values in (v) and again $v_1$ and
$\rho$ for $F_{j}^{n+1}$ and the values in (iii) for $F_{j}^{n}$.
\item[vii.] Test the calculations in (vi) for accuracy and convergence. If they are
sufficiently accurate, move to the next time step, otherwise repeat
steps v. and vi. with the new values of $v_1$ and $\rho$.
\end{itemize}

As with the standard characteristic model, described in
\cite{bishop96, vanderwalt10}, calculations in the regions around
$\lambda=0$ require special consideration. The mechanisms used
in \cite{vanderwalt10} for these regions was directly adapted for
the affine CIVP model. In the $\lambda \approx 0$ region, it is is
evident from the occurrence of $\lambda$ denominators that equations
(\ref{sec:affine.v1u}) and (\ref{sec:affine.rhou}) will not be
well-behaved. Consequently, the $\lambda \approx 0$ region is
calculated by making use of second order series expansions. The
region where the series solution meets with the CIVP solution, also
requires special treatment to avoid artificial instabilities. This
has been done by smoothing out the merger region with a weighted
average between the two solutions.

\section{Transformations: Comoving coordinates}
\label{sec:affine.transform}

\subsection{General transformation}

As a measure of the degree of convergence and accuracy, numerical
calculations have to be compared with results from known solutions,
which require coordinate transformations from conventional comoving
cosmological coordinates to affine null coordinates. These
transformations follow directly from calculating the geodesic paths
in space and time. This has the general form
\begin{align}
\frac{d^2x^a}{d\lambda^2}
    + \Gamma _{bc} ^{a} \frac{dx^b}{d\lambda} \frac{dx^c}{d\lambda}
    = 0 . \label{sec:tran.geo}
\end{align}

The parabolic LTB model, which is spherically symmetric and radially
inhomogeneous will be used as the general case for transformations.
Its geometry in comoving synchronous coordinates is described by the
metric \footnote{Note that $r$ here is the comoving radial distance
and not the diameter distance.}
\begin{align}
ds^2 = -dt^2
    + [R_{,r}(t,r)]^{2}dr^2 + [R(t,r)]^2\{d\theta ^2 + \sin ^2 \theta d\varphi
    ^2\} .
    \label{sec:ltb.met}
\end{align}
Here, $t$ is the cosmic (proper) time, $r$ is a comoving radial
coordinate with $\theta$ and $\varphi$ the inclination and azimuth
angles. $R(t,r)$ is the areal radius and $4\pi R^2$ defines the
proper surface area of a sphere with coordinate radius $r$ at a
constant time slice \cite{ellis98}.

Using (\ref{sec:tran.geo}), the geodesic equations for the LTB model
becomes
\begin{align}
\frac{d^2t}{d\lambda^2}
    + \Gamma_{11}^{0} \left(\frac{dr}{d\lambda}\right)^2 &=0 \\
\frac{d^2r}{d\lambda^2}
    + 2 \, \Gamma_{01}^{1} \frac{dt}{d\lambda} \frac{dr}{d\lambda}
    + \Gamma_{11}^{1} \left(\frac{dr}{d\lambda}\right)^2
    &= 0.
\end{align}
When scaled to some maximum time, $t_0$, as the current age of a
universe, the conditions at $\lambda=0$ are $t = t_{0}$ and $r = 0$
with the initial directions constrained to be null by
${dt}/{d\lambda} = 1$ and ${dr}/{d\lambda} = R(t_{0},0)$. As
comparative values on a null cone grid, the covariant velocity
follows directly from $v_1=dt/d\lambda$, the diameter distance from
$\hat{r}(\lambda) = R(t,r(\lambda))$, while $\rho$ is determined
from the coordinate expression for the specific model using
$(t(\lambda),r(\lambda))$ as the coordinates for $t$ and $r$ on a
null cone.

\subsection{Models for code verification}
\label{sec:ver.models}

\subsubsection{Einstein-de Sitter model}
\label{sec:ver.eds}

As an illustration of comoving to affine transformations, the
Einstein-de Sitter model (EdS), scaled to $t_0=1$ with $G=1$, which
has the metric
\begin{align}
ds^2 = -dt^2 + t^{4/3} dr^2 + t^{4/3}r^2\{d \theta^2 + \sin^2 \theta
d \varphi^2\} .
\end{align}
provides a system of equations from which $t(\lambda)$ and
$r(\lambda)$ can be obtained
\begin{align}
\frac{d^2t}{d\lambda^2}
    + \frac{2}{3} t^{1/3} \left(\frac{dr}{d\lambda}\right)^2 &=0 \\
\frac{d^2r}{d\lambda^2}
    + \frac{4}{3} t^{-1} \frac{dt}{d\lambda} \frac{dr}{d\lambda} &=
    0.
\end{align}
At $\lambda=0$: $t = 1$, $r = 0$, ${dt}/{d\lambda} = -1$ and
${dr}/{d\lambda} = t^{-2/3}$.

\begin{figure}[h!]
\begin{center}$
\begin{array}{l}
\vspace{-30pt} \hspace{-40pt}\includegraphics[width=0.37\textwidth,
angle=-90]{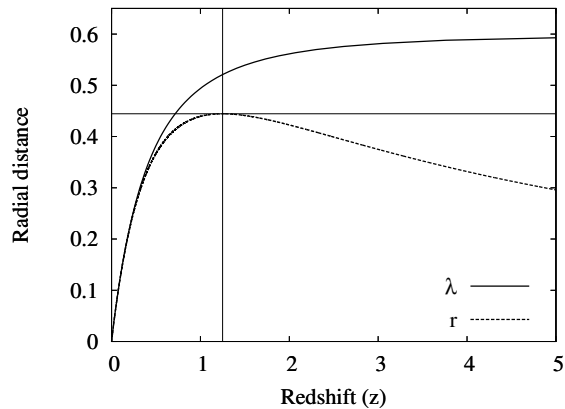}
\end{array}$
\end{center} \vspace{20pt} \caption{The radial distance coordinates $d_D$ and $\lambda$
related to $z$, which is a common observational measure. }
\label{sec:affine.z-l}
\end{figure}

Solving these equations numerically provides a useful example of the
motivation for working with an affinely parameterized radial
coordinate as opposed to the diameter distance. This is illustrated
in figure \ref{sec:affine.z-l} where the diameter distance and
affine parameter is plotted against the redshift. In terms of
observations, with the emphasis on the AH at $z=1.25$, the diameter
distance reaches its maximum and then desreases while $\lambda$
keeps on increasing and provides a unique coordinate for higher
redshifts.

Substituting $t(\lambda)$ into the density equation $\rho=1/(6\pi
t^2)$ provides the density profile on a specific null cone. It is
interesting to point out here that as $t$ varies with $\lambda$ on
each null cone, $\rho$ also varies i.e. even though the Einstein-de
Sitter model is homogeneous in conventional cosmological
coordinates, null cones are not hypersurfaces of radial homogeneity
and the model becomes inhomogeneous in null coordinates (see
\cite{Ribeiro95}).

\subsubsection{$\Lambda$CDM}

The $\Lambda$CDM or concordance model is of interest for the current
understanding of the Universe and is also useful to test the affine
CIVP model with a non-zero cosmological constant. In terms of the
LTB metric (\ref{sec:ltb.met}), using cosmological properties at the
current epoch ($t_0$), the solution of the flat $\Lambda$CDM model
is (see \cite{lidsey09}):
\begin{align}
R(t,r) = \left( \frac{\Omega_{m0}}{\Omega_{\Lambda}}\right)^{1/3}
\left(\sinh\left[ \frac{3}{2} H_0 \sqrt{\Omega_{\Lambda}} \,
t\right] \right)^{2/3} r
\end{align}
with the age of the Universe:
\begin{align}
t_0 = \frac{2}{3} \left(H_{0} \sqrt{\Omega_{\Lambda}}\right)^{-1}
\sinh^{-1}\left[
\left(\frac{\Omega_{\Lambda}}{\Omega_{m0}}\right)^{1/2} \right]
\end{align}
and the density distribution
\begin{align}
\rho_m = \frac{3 H_0^2}{8 \pi G}  \frac{\Omega_{m0} R_0^3}{R^3}.
\end{align}
Here, $\Omega_{m0}$ and $\Omega_{\Lambda}$ are the current density
parameters for Baryonic matter and the cosmological constant
respectively and $H_0$ is the current Hubble constant. Values
representative of the actual Universe are: $\Omega_{m0}=0.3$,
$\Omega_{\Lambda}=0.7$ and $H_0=72$ Mpc s$^{-1}$ km$^{-1}$. For the
purpose of code verification, however, using geometric units ($G=1$)
and rescaling time to have $t_0=1$, a dimensionless value of $H_0 =
0.964$ was used to make comparison with other models more
convenient.

\subsubsection{Lema\^{i}tre-Tolman-Bondi}

LTB models, introduced in \cite{lemaitre33,tolman34,bondi47}, have
been popular in recent years to demonstrate how inhomogeneities can
reproduce the effects of type Ia supernovae redshift dimming without
the need of a cosmological constant (see
\cite{celerier00,pas99,enqv08,gar08,celerier07} amongst others). It,
therefore, has physical significance and also provides a general
case for testing the effect of radial inhomogeneity on the null cone
behaviour. The test cases investigated, here, are for simplicity
restricted to the parabolic solution. Substituting
(\ref{sec:ltb.met}) into the EFEs and solving gives (see
\cite{krasinski06,musta98.2}):
\begin{align}
R(t,r) = \left[\frac{9}{2} M(r) (t-t_{B}(r))^{2}\right]^{1/3}
\textrm{~~and~~} \rho(t,r) = \frac{M_{,r}}{4 \pi \; R^2 R_{,r}}
\textrm{ .} \label{sec:ltb-e0}
\end{align}
$M(r)$ is the \emph{active gravitational mass} which is the mass
contributing to the gravitational field and $t_B(r)$ is defined as
the \emph{bang time function}, which is a surface defined by the
local time at which $R=0$. A simplified subset of models with
mathematical convenient properties follow by setting $M(r) =
M_0=r^3$ as a coordinate condition where $M_0$ is a constant which
for illustrative purposes is set to $2/9$. Equations
(\ref{sec:ltb-e0}) then reduce to:
\begin{align}
R(t,r) &= r (t-t_{B}(r))^{2/3} \label{sec:civp-R}
\end{align}
with $\rho$:
\begin{align}
\rho(t,r) &= \frac{1}{2 \pi (t-t_B(r))(3t - 3t_B(r)-2rt_{B,r}(r))} .
\label{sec:civp-ltbrho}
\end{align}

By selecting $t_B(r)=0$, equation (\ref{sec:civp-ltbrho}) becomes
the EdS model, which can be verified against the results in section
\ref{sec:ver.eds}. If $t_B(r) \neq 0$ and $t_{B,r}(r)=0$, the time
of the initial singularity is adjusted and the age of the Universe
changes. For non-constant functions, the initial singularity becomes
a singular surface and the age of the Universe becomes subject to
the position of an observer (i.e. the age of the Universe depends on
$r$). Thus, a variety of models can be generated for testing the
code on parabolic spatial sections.

A simple choice of bang function is implemented as a verification
model:
\begin{align}
t_B(r) = b r , \label{sec:civp-tB}
\end{align}
with $b$ being a constant. This simplifies equations
(\ref{sec:civp-R}) and (\ref{sec:civp-ltbrho}) to:
\begin{align}
R(t,r) &= r (t-b r)^{2/3} \label{sec:civp-Rkx} \\
\rho(t,r) &= \frac{1}{2 \pi (t-br)(3t -5br))} .
\label{sec:civp-ltbrhok}
\end{align}

By varying the value of $b$, different aspects of inhomogeneity can
be studied. The value $b=0$ is exactly the EdS model and $b>0$
shifts the age of a universe to a younger age as $r$ increases while
$b<0$ provides the opposite effect where a universe is shifted to an
older age as $r$ increases. The latter case is particularly
interesting since it provides a mechanism to mimic a cosmological
constant on low redshifts \cite{celerier00}. This is, however, not
necessarily the case for high redshifts and the case where $b=-0.5$
is used as a model where the effect of inhomogeneity is clearly
visible on higher redshifts.

\clearpage
\section{Results}
\label{sec:affine.results}

\subsection{Verification results}

The results of numerical calculations of the models described in
section \ref{sec:ver.models} are presented in this section. It is
shown how the diameter distance ($\hat{r}$), the covariant velocity
($v_1$) and the density $\rho$ evolve on PNCs in the past from
observations on the current PNC. Also of interest is the evolution
of $\hat{r}$ against the redshift $z$, which provides an
illustration of an observer's perception of distance from a
particular PNC vertex. In the results the current PNC values are
indicated with $u_0$, the oldest PNC with $u_{\max}$ and an
intermediate PNC is also shown between these values. Since the
purpose of the calculations is to test the behaviour of the
numerical model and not the physics of the specific cosmological
models, time is scaled to $u_0=1$ and geometric coordinates are used
to provide dimensionless results. In all cases the numerical results
(points in the figures) closely follow the transformed results
(lines in the figures). Table \ref{sec:res.tab.sum} gives a summary
of the test cases.

\begin{table}[h!]
\begin{center}
\begin{tabular}{ l c c c c c}
\hline
Model & ~~$z$ at $r_{max}$ & ~~$z_{max}$~~ & ~~$u_{0}$~~ & ~~$u_{max}$~~ & ~~Figures \\
\hline
EdS & $1.25$ & $4.5$ & $1$ & $0.35$ & \ref{sec:res.eds.r}, \ref{sec:res.eds.rhov} \\
$\Lambda$CDM ($\Omega_{\Lambda}=0.7$) \footnote{This result is in agreement with that of \cite{araujo09.1}} & $1.61$ & $6$ & $1$ & $0.4$ & \ref{sec:res.cmd.r}, \ref{sec:res.cmd.rhov} \\
LTB ($b=-0.5$) & $1.02$ & $1.2$ & $1$ & $0.2$ & \ref{sec:res.ltb.r}, \ref{sec:res.ltb.rhov}\\
\hline
\end{tabular}
\end{center}
\caption{Affine CIVP test cases.} \label{sec:res.tab.sum}
\end{table}

\begin{figure}[h!]
\begin{center}$
\begin{array}{ll}
\hspace{-25pt}
\includegraphics[width=0.37\textwidth, angle=-90]{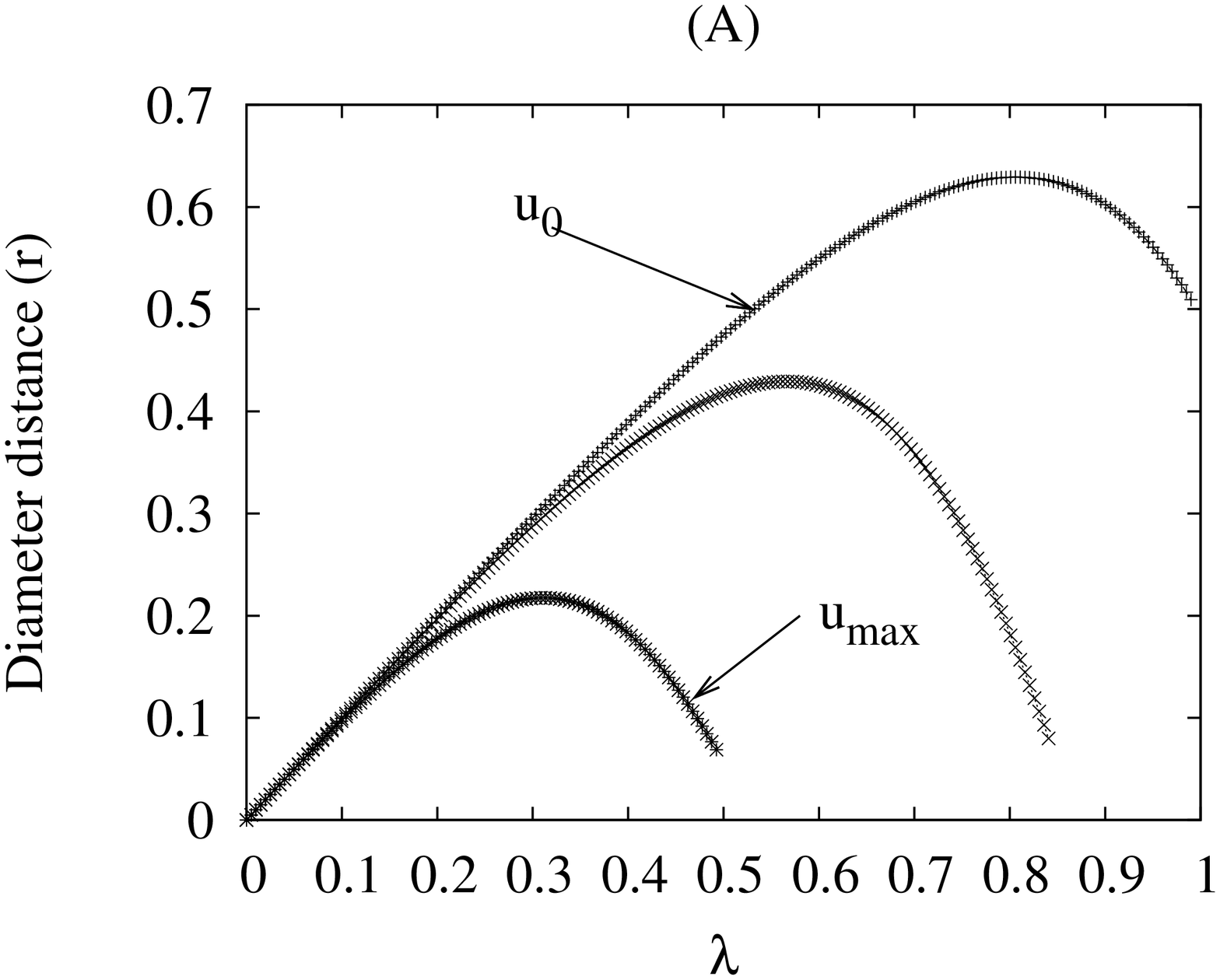} &
\hspace{-30pt}\includegraphics[width=0.37\textwidth,
angle=-90]{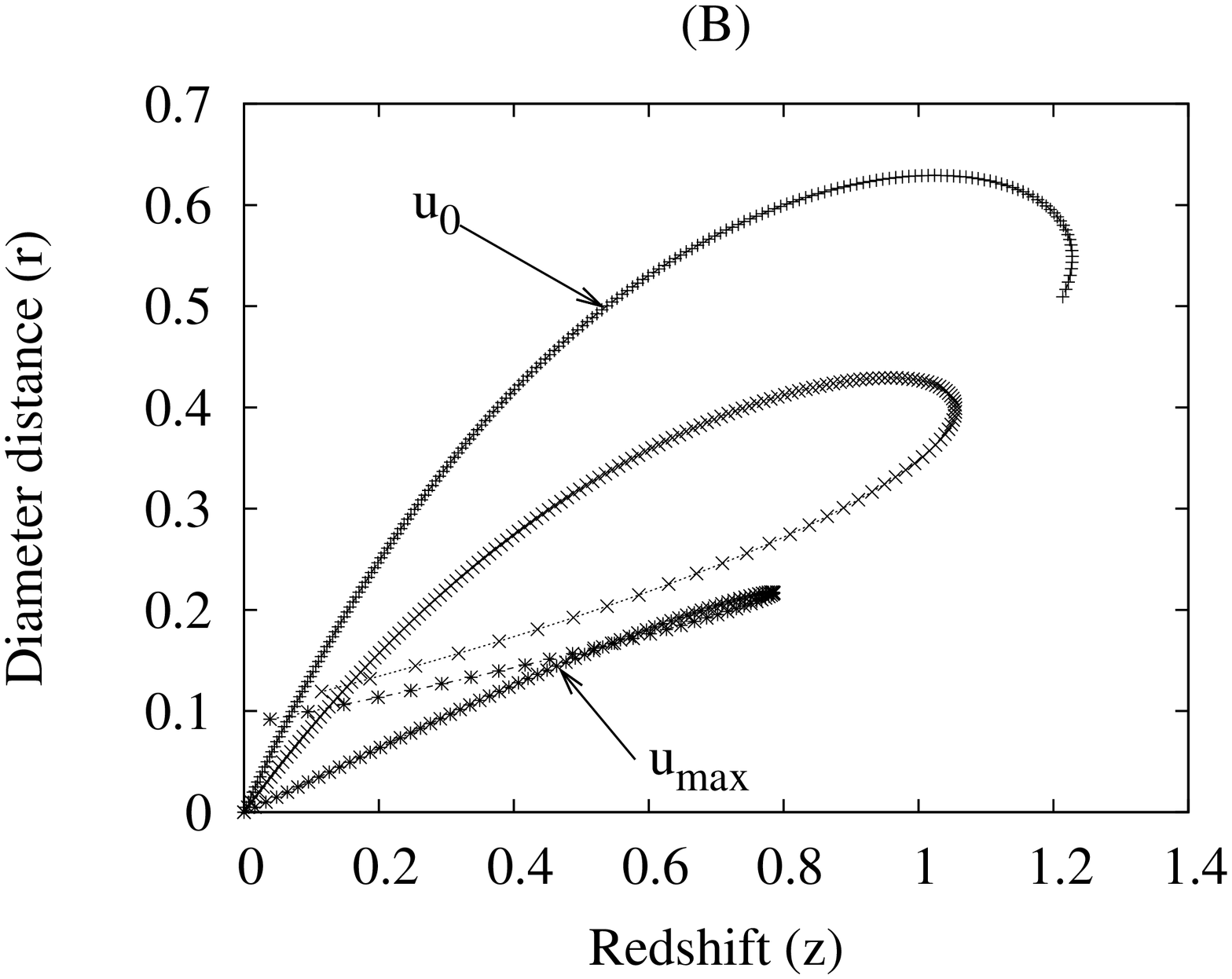}
\end{array}$
\end{center} \vspace{-10pt} \caption{Diameter distance against $\lambda$ (A) and
against of $z$ (B) on PNCs at different proper times ($u$) evolved
from a local PNC up to $z=1.2$.} \label{sec:res.ltb.r}
\end{figure}

Figure \ref{sec:res.ltb.r} B is a particularly interesting
illustration of an observer's perception of distance in an
inhomogeneous universe. Here, the $(\hat{r},z)$ behaviour on earlier
PNCs includes loops, which will completely obscure the perception of
diameter distance and redshift as measures of distance. This type of
behaviour provides insight into the physical nature of a model
by investigating its past behaviour. Although the initial PNC of the
specific LTB model is not significantly out of line compared to more
accepted models, especially on low redshifts, its past behaviour
is rather unusual. This does not make it unrealistic, but does say
that an observer at $u=u_{\max}$ would have found the interpretation
of cosmological data particularly difficult.

\subsubsection{Einstein-de Sitter}

\vspace{-20pt}

\begin{figure}[h!]
\begin{center}$
\begin{array}{ll}
\hspace{-25pt}
\includegraphics[width=0.37\textwidth, angle=-90]{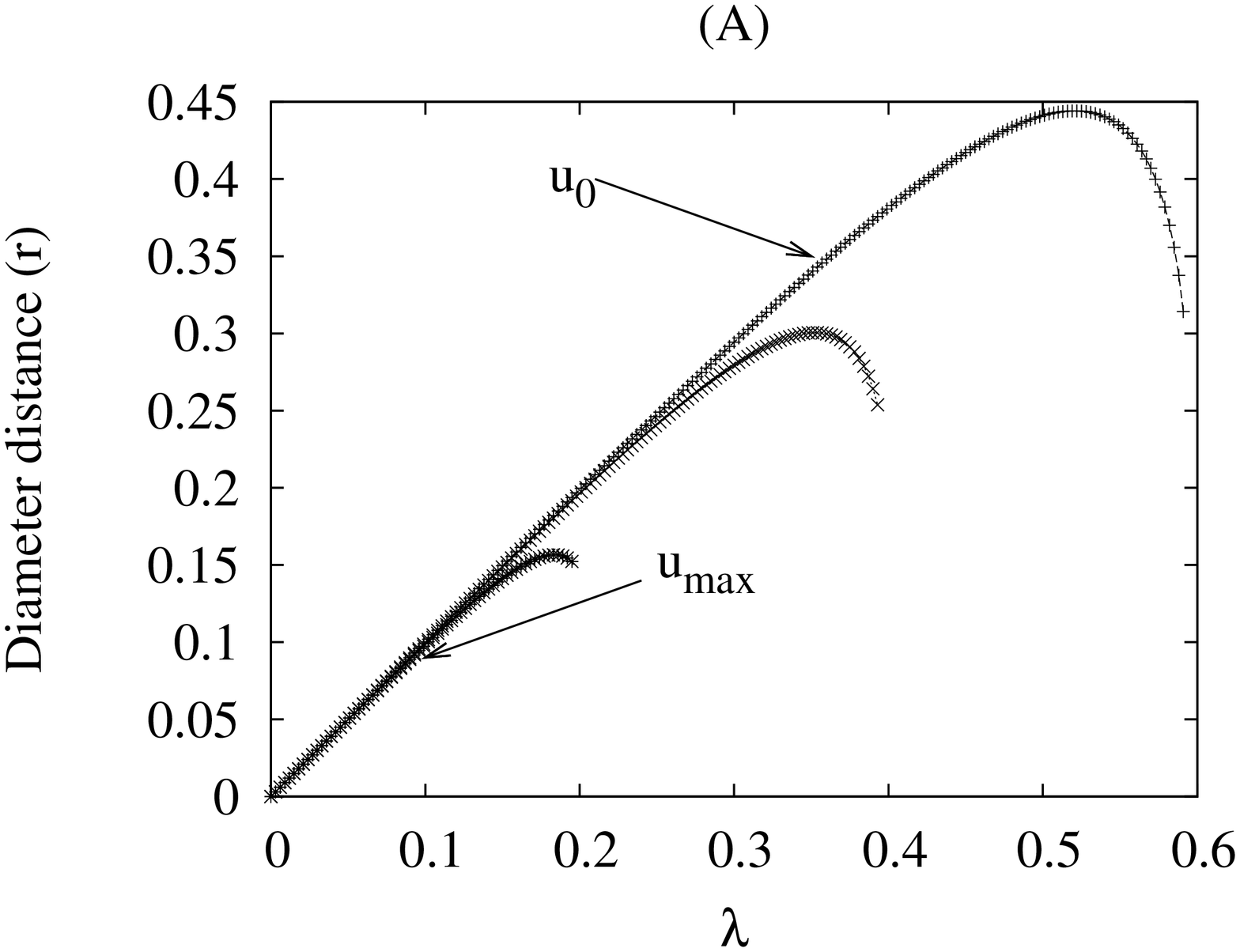} &
\hspace{-30pt}\includegraphics[width=0.37\textwidth,
angle=-90]{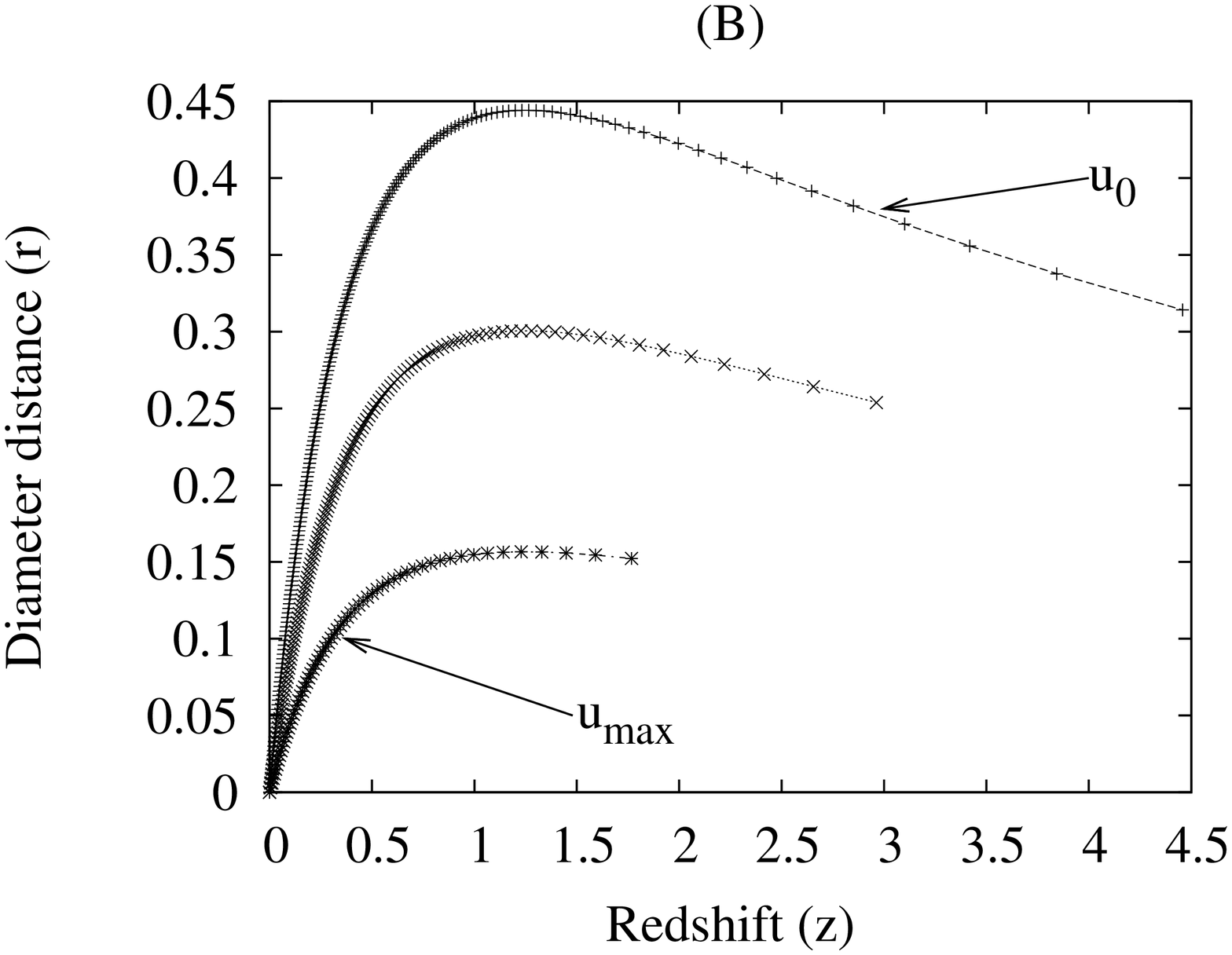}
\end{array}$
\end{center} \vspace{-10pt} \caption{Diameter distance against $\lambda$ (A) and
against of $z$ (B) on PNCs at different proper times ($u$) evolved
from a local PNC up to $z=4.5$.} \label{sec:res.eds.r}
\end{figure}

\begin{figure}[h!]
\begin{center}$
\begin{array}{ll}
\hspace{-25pt}
\includegraphics[width=0.37\textwidth, angle=-90]{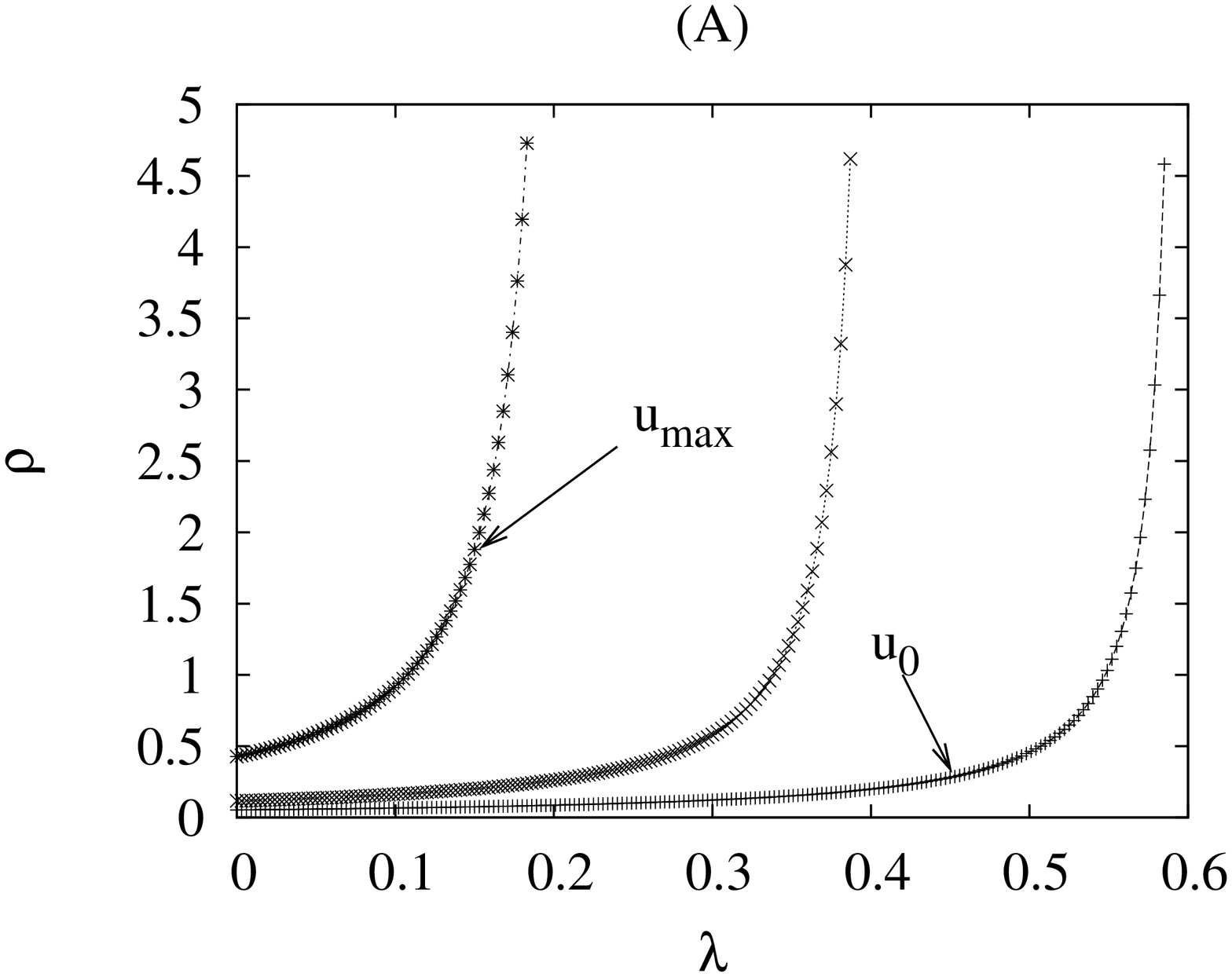} &
\hspace{-30pt}\includegraphics[width=0.37\textwidth,
angle=-90]{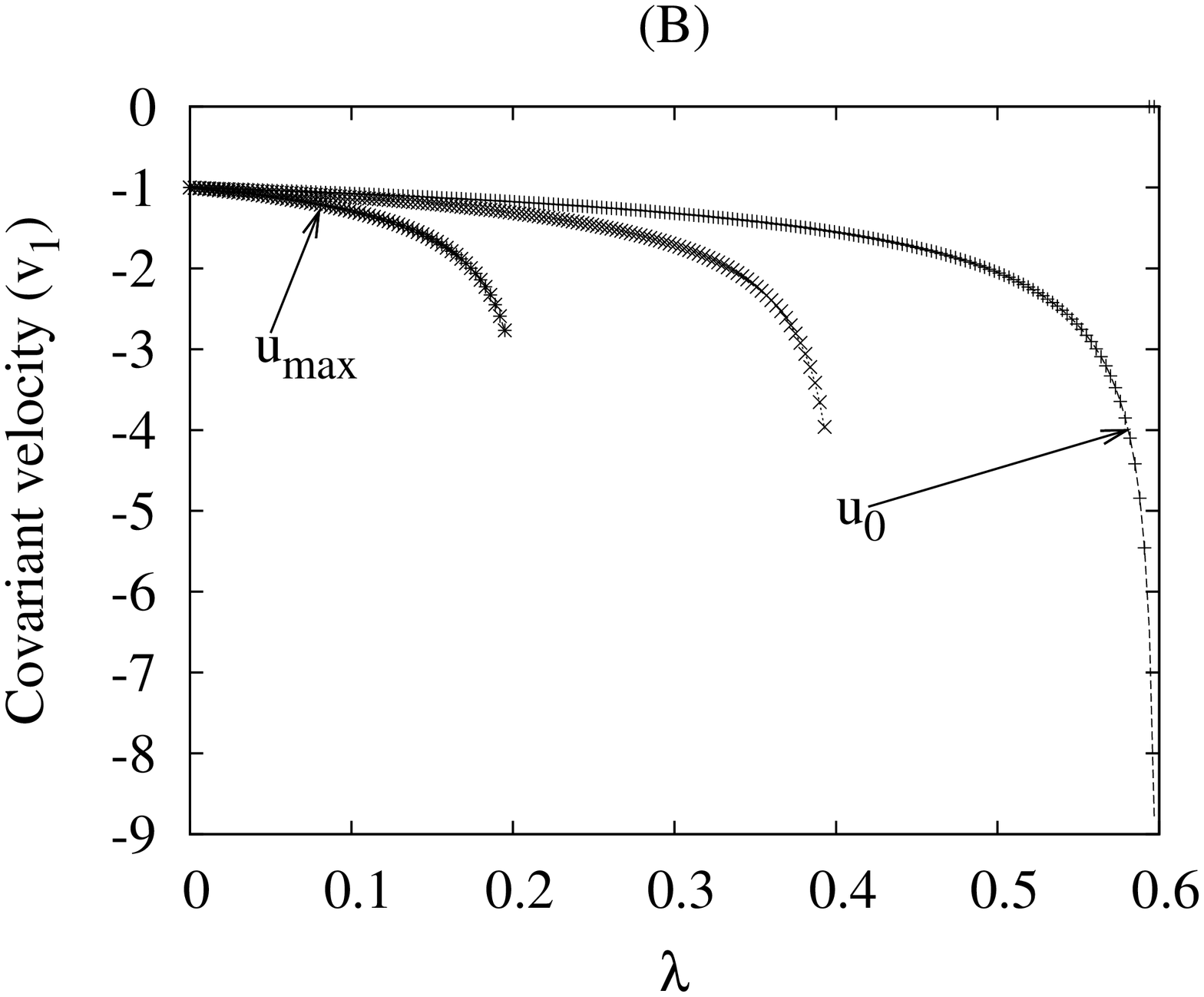}
\end{array}$
\end{center} \vspace{-10pt} \caption{Density distribution (A) and covariant velocity
(B) on PNCs at different proper times ($u$) evolved from a local PNC
up to $z=4.5$.} \label{sec:res.eds.rhov}
\end{figure}

\newpage

\subsubsection{$\Lambda$CDM with $\Omega_{\Lambda}=0.7$}

\vspace{-10pt}

\begin{figure}[h!]
\begin{center}$
\begin{array}{ll}
\hspace{-25pt}
\includegraphics[width=0.37\textwidth, angle=-90]{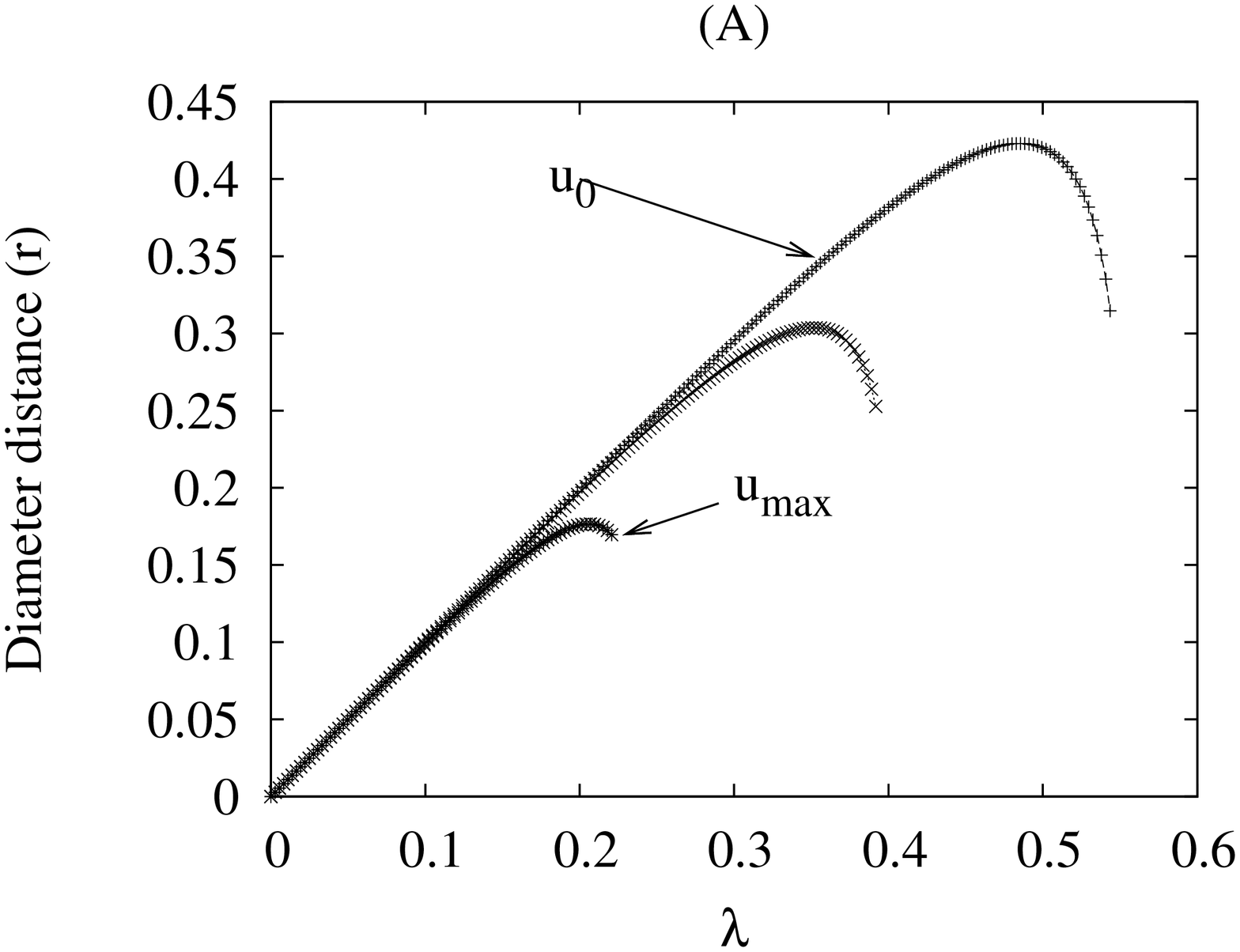} &
\hspace{-30pt}\includegraphics[width=0.37\textwidth,
angle=-90]{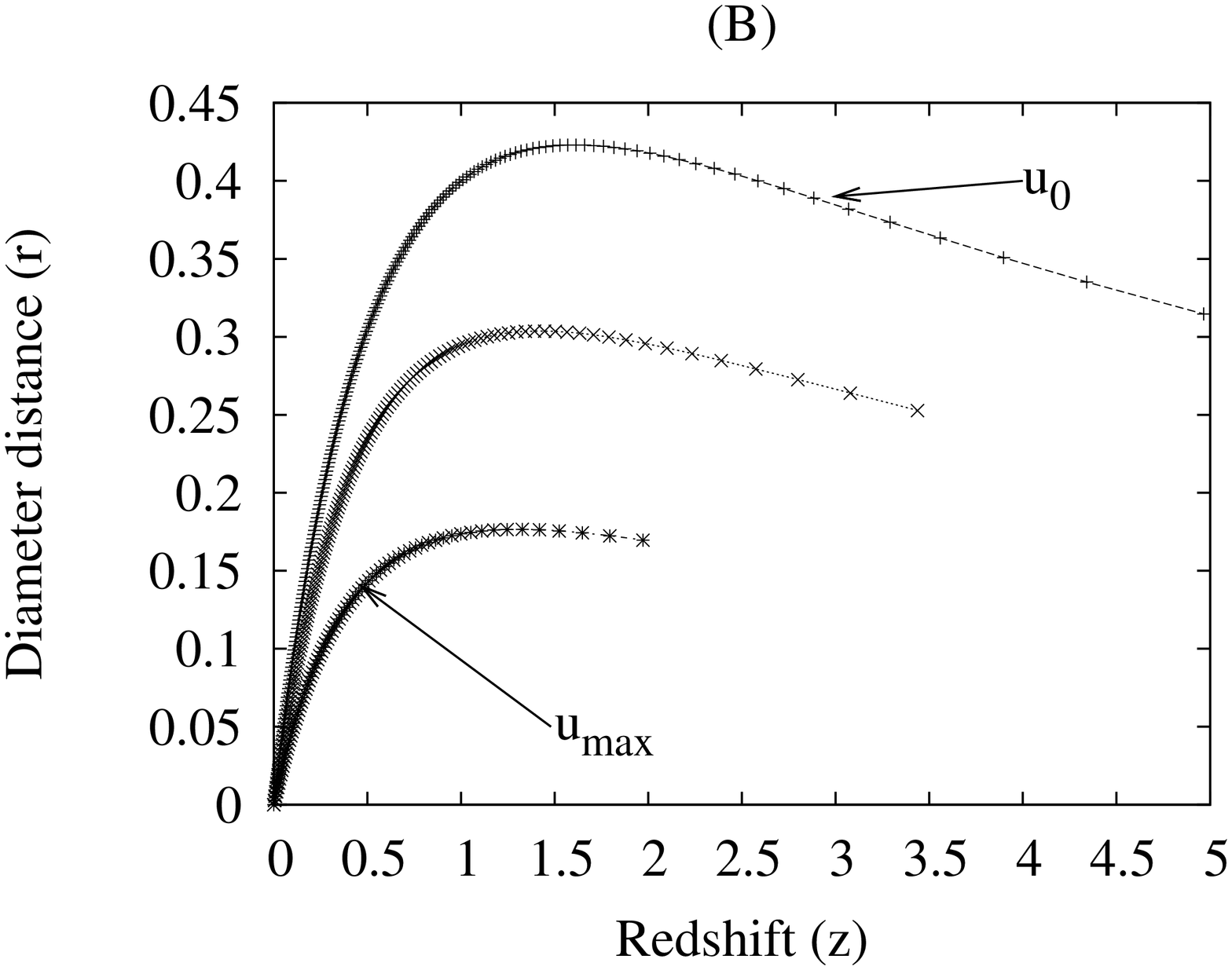}
\end{array}$
\end{center} \vspace{-10pt} \caption{Diameter distance against $\lambda$ (A) and
against of $z$ (B) on PNCs at different proper times ($u$) evolved
from a local PNC up to $z=5$.} \label{sec:res.cmd.r}
\end{figure}

\begin{figure}[h!]
\begin{center}$
\begin{array}{ll}
\hspace{-25pt}
\includegraphics[width=0.37\textwidth, angle=-90]{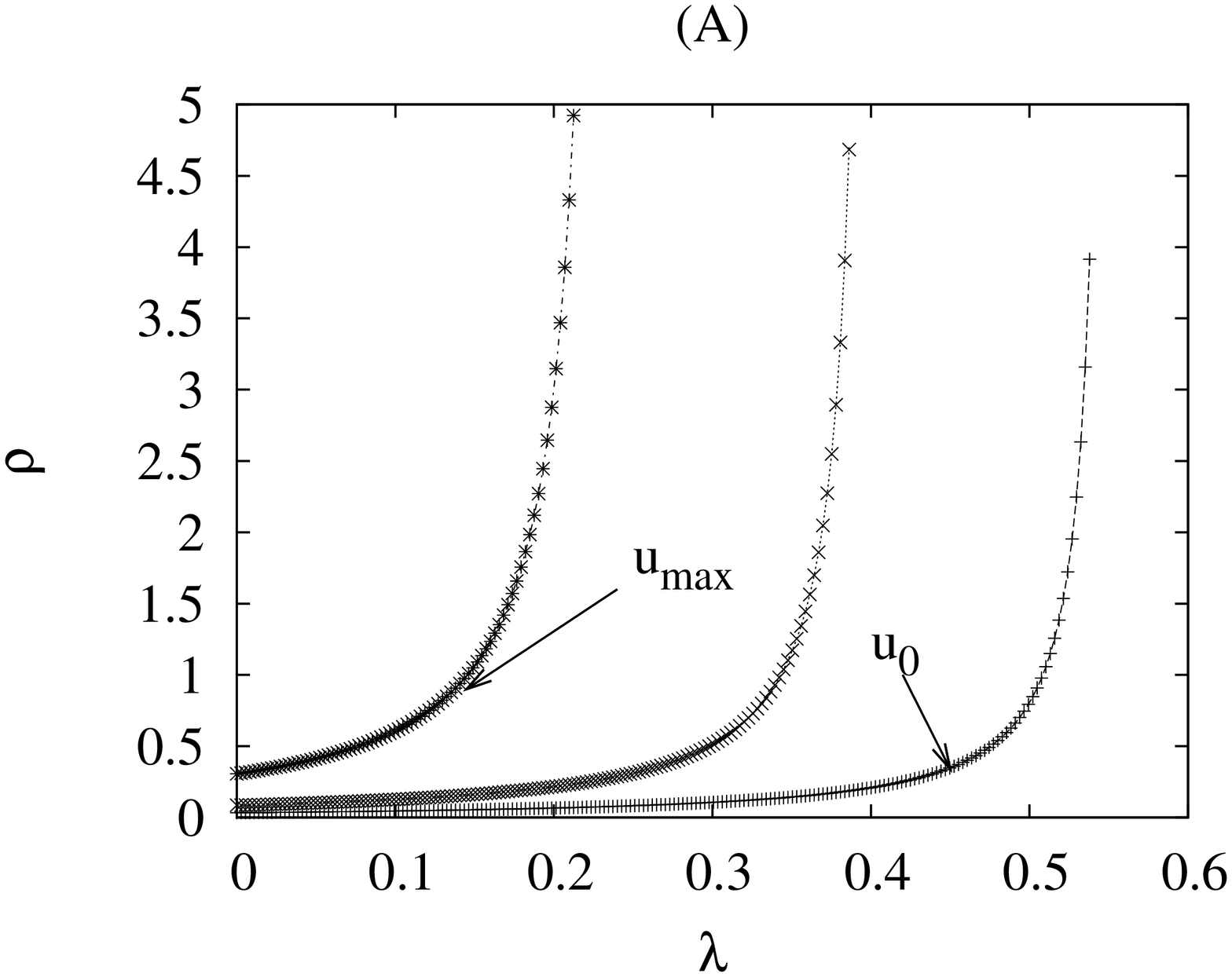} &
\hspace{-30pt}\includegraphics[width=0.37\textwidth,
angle=-90]{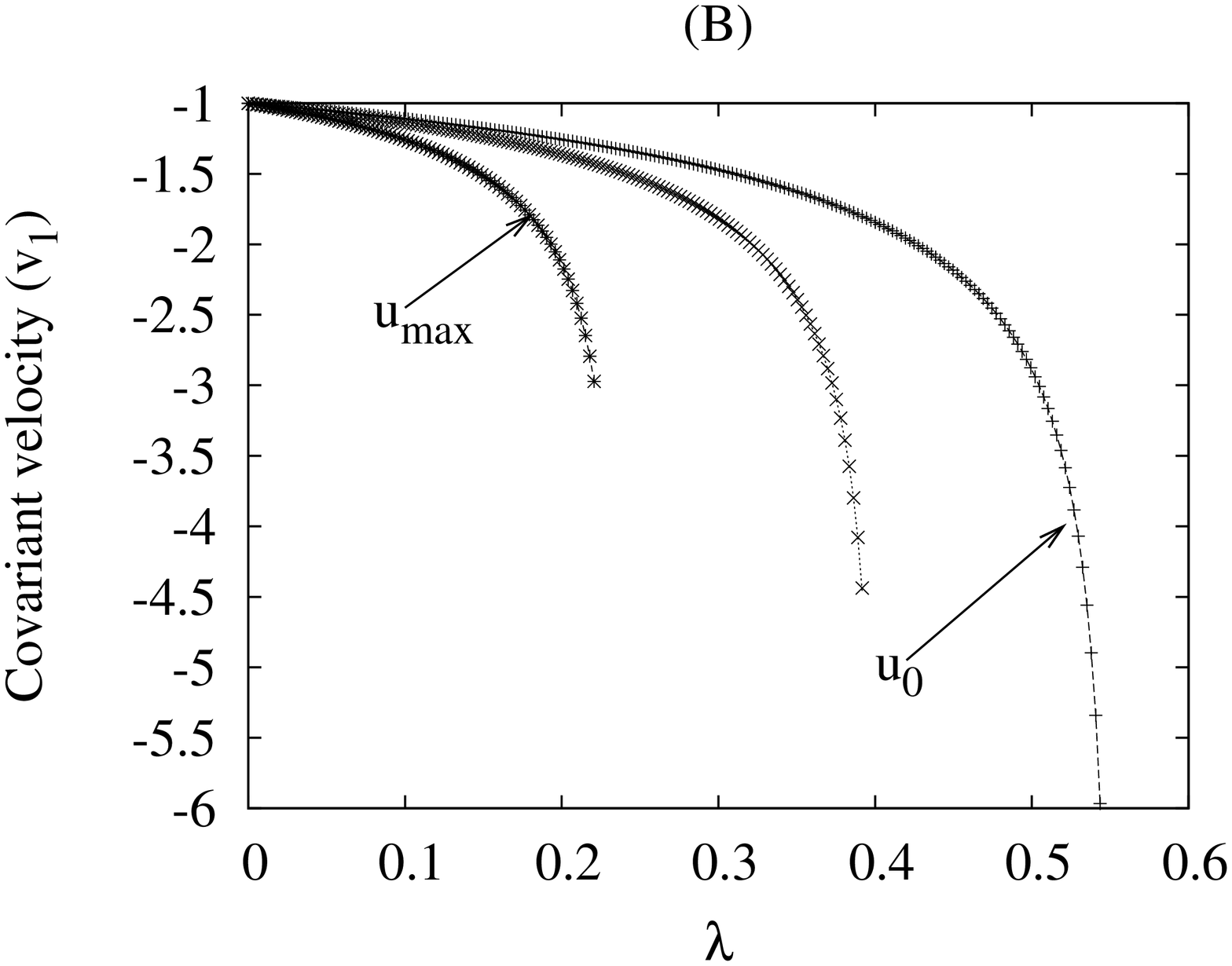}
\end{array}$
\end{center} \vspace{-10pt} \caption{Density distribution (A) and covariant velocity
(B) on PNCs at different proper times ($u$) evolved from a local PNC
up to $z=5$.} \label{sec:res.cmd.rhov}
\end{figure}

\newpage

\subsubsection{LTB with $b=-0.5$}

\vspace{-10pt}

\begin{figure}[h!]
\begin{center}$
\begin{array}{ll}
\hspace{-25pt}
\includegraphics[width=0.37\textwidth, angle=-90]{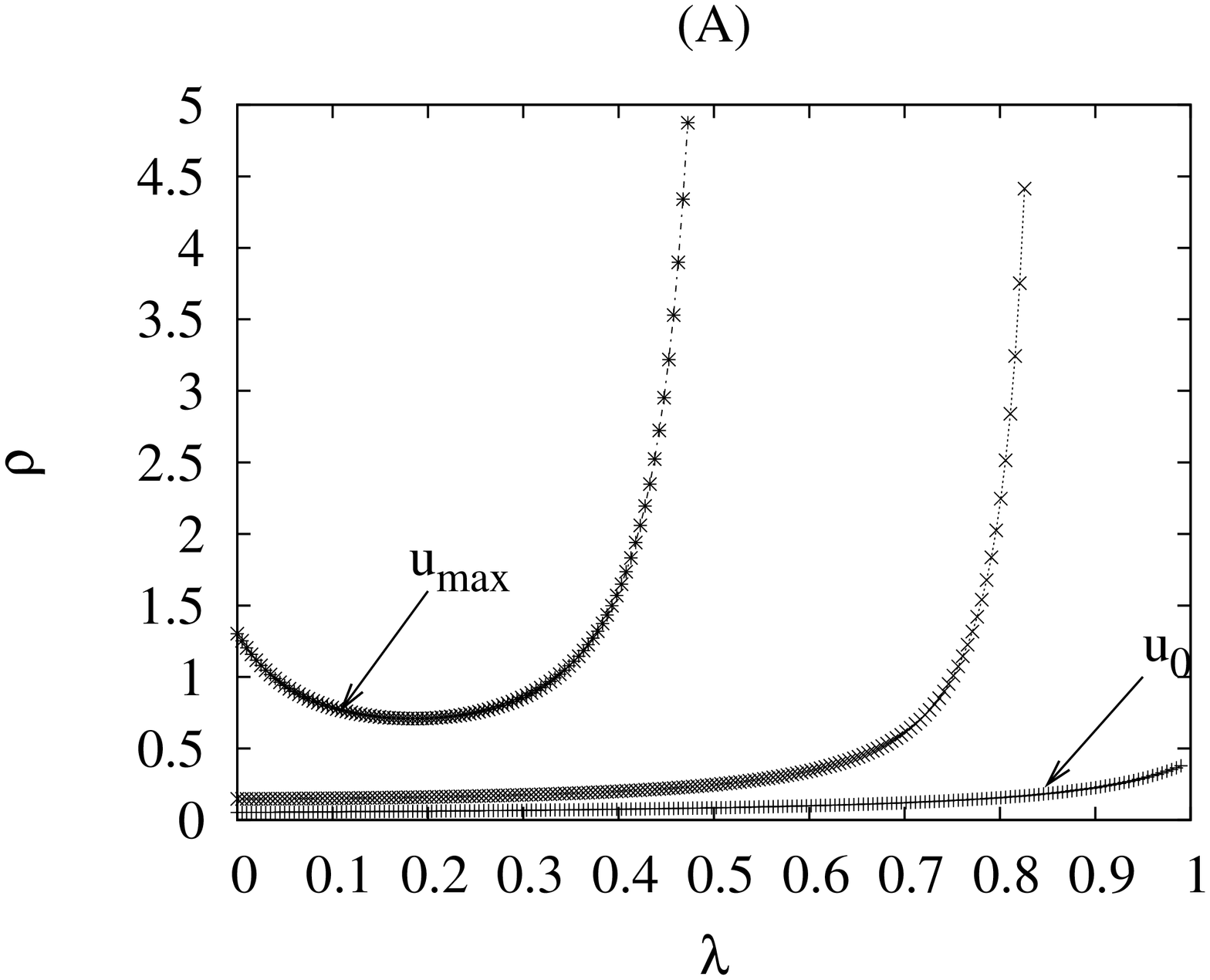} &
\hspace{-30pt}\includegraphics[width=0.37\textwidth,
angle=-90]{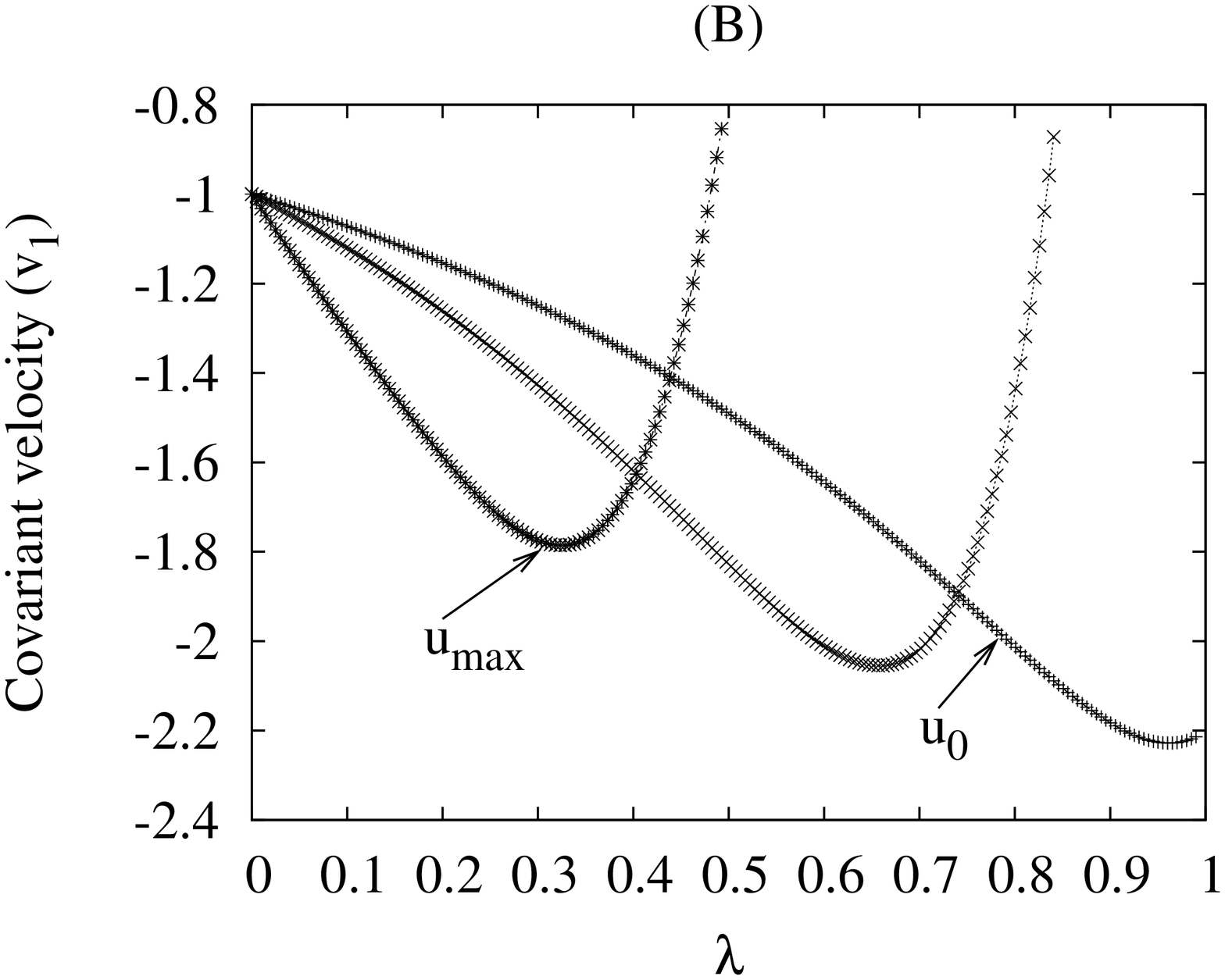}
\end{array}$
\end{center} \vspace{-10pt} \caption{Density distribution (A) and covariant velocity
(B) on PNCs at different proper times ($u$) evolved from a local PNC
up to $z=1.2$.} \label{sec:res.ltb.rhov}
\end{figure}

\newpage
\subsection{Convergence and accuracy}

\begin{figure}[h!]
\begin{center}$
\begin{array}{ll}
\hspace{-60pt}
\includegraphics[width=0.37\textwidth, angle=-90]{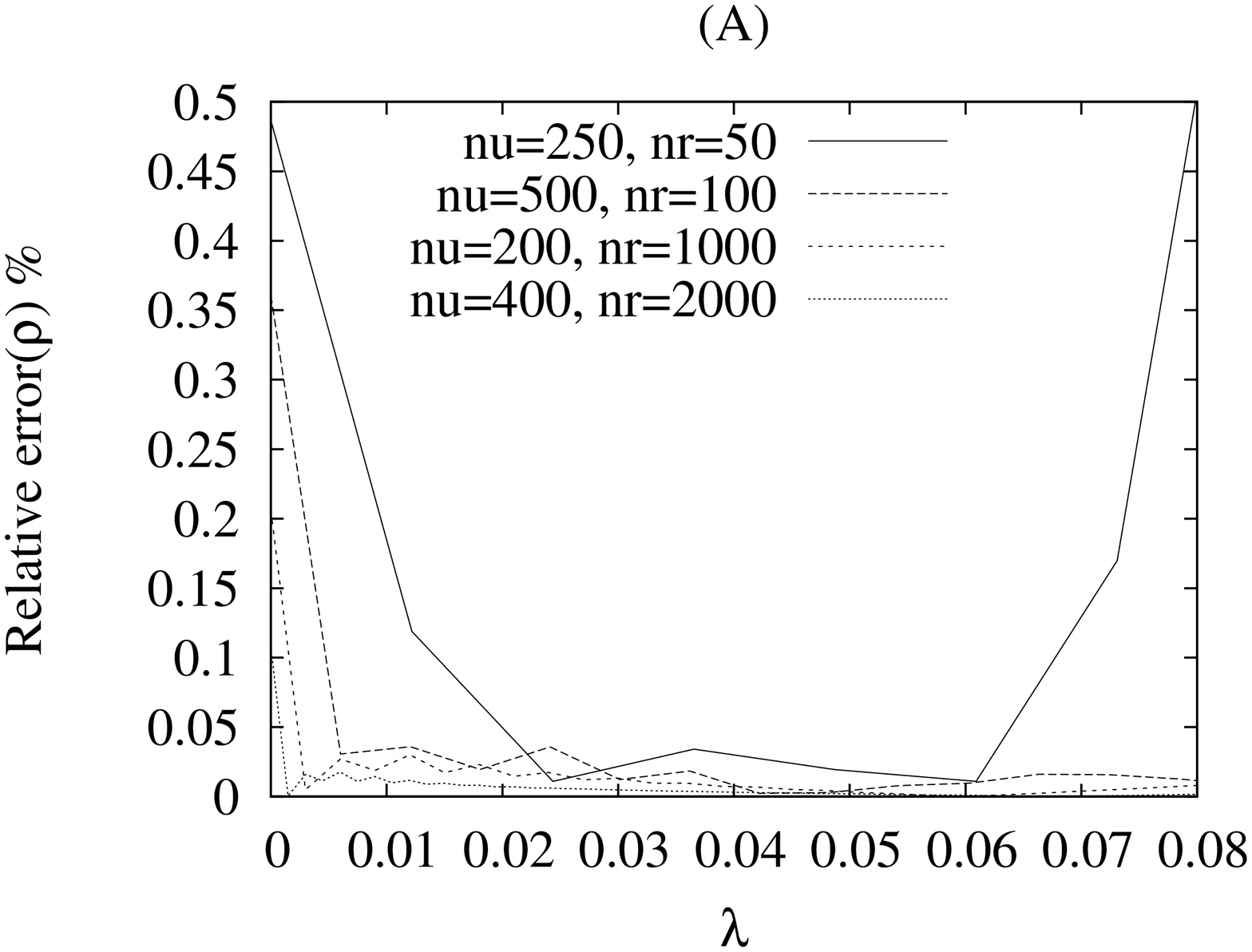} &
\hspace{-10pt}\includegraphics[width=0.37\textwidth,
angle=-90]{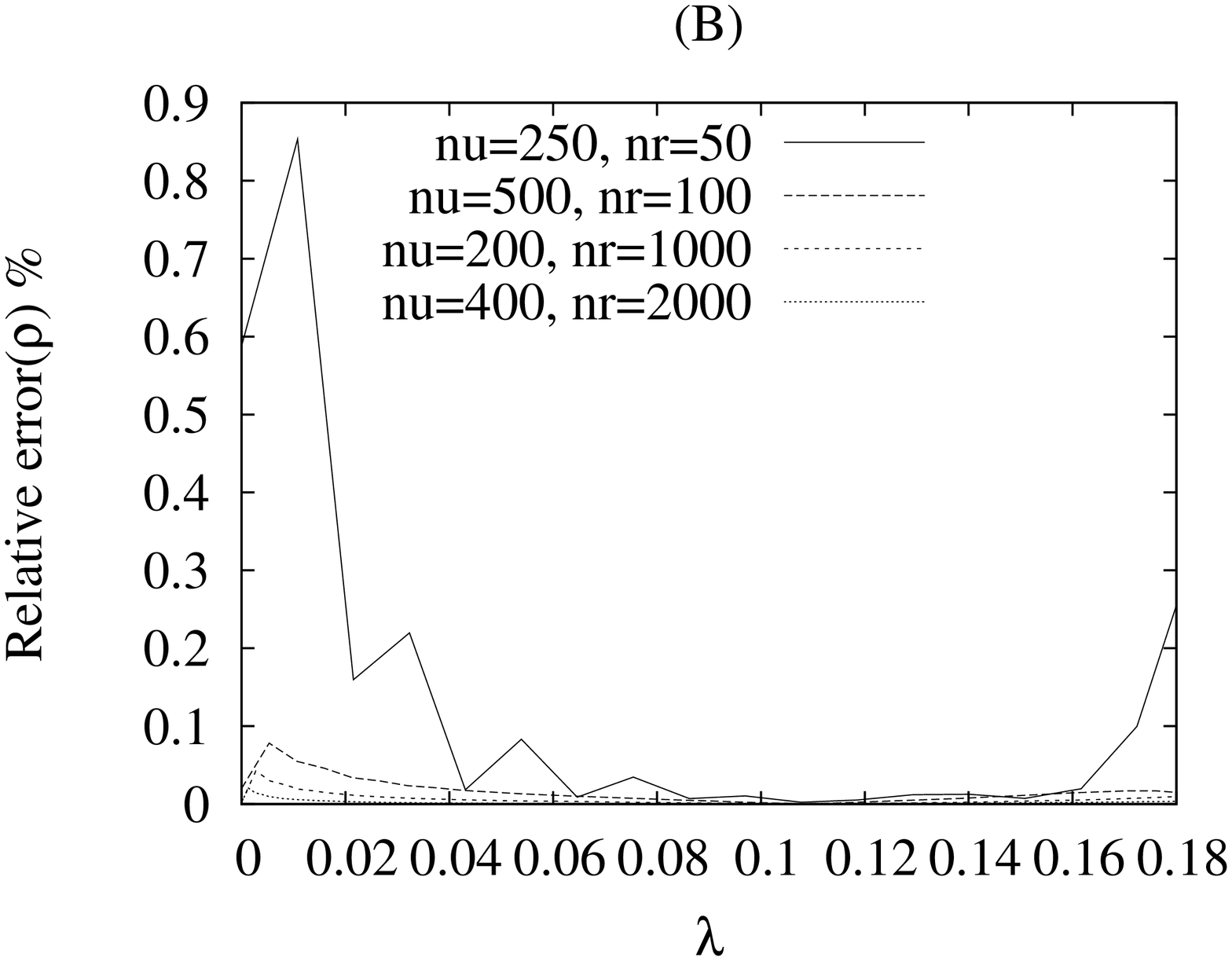} \\
\hspace{-60pt}
\includegraphics[width=0.37\textwidth, angle=-90]{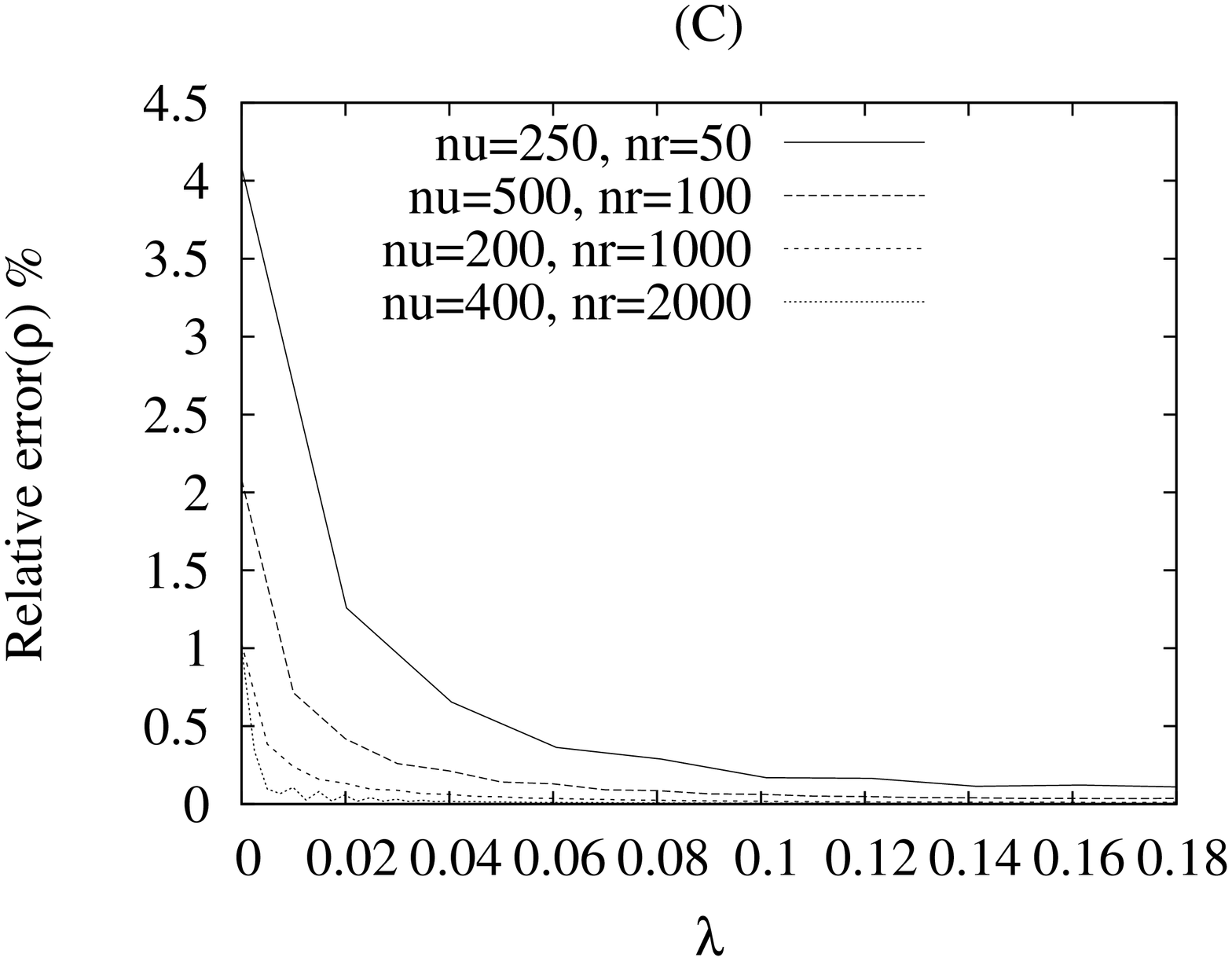} &
\hspace{-10pt}\includegraphics[width=0.37\textwidth,
angle=-90]{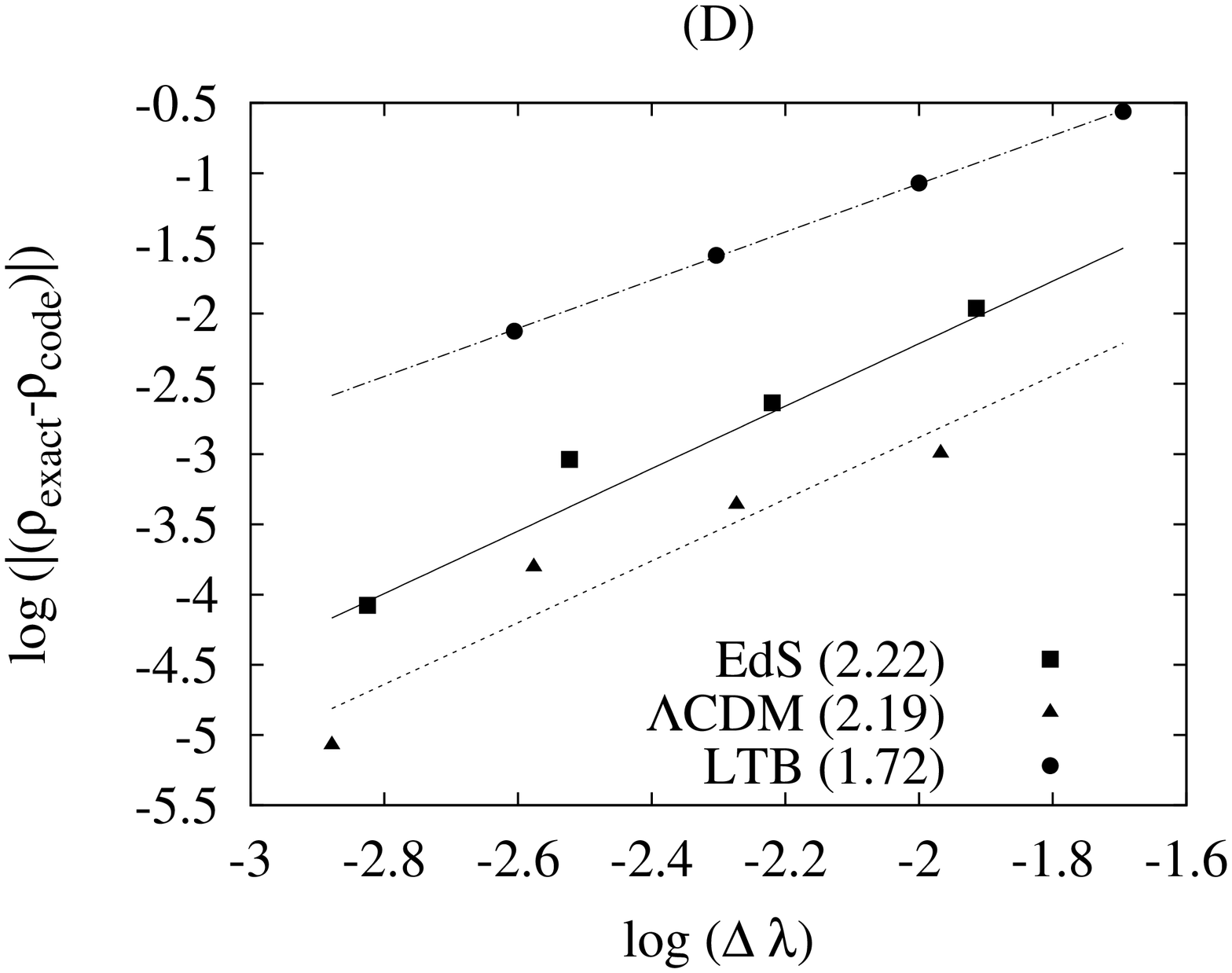}
\end{array}$
\end{center} \vspace{-10pt} \caption{Radial error distribution on the oldest null cones
for the Einstein-de Sitter ($u_{max}=0.35$)(A), $\Lambda$CDM
($u_{max}=0.4$)(B) and LTB ($u_{max}=0.2$)(C) models for different
grid resolutions. In (D), the error against grid size is shown; the
slope of the change in error represents the order of convergence and
is indicated in brackets in the legend. } \label{sec:res.con.acc}
\end{figure}

Figures \ref{sec:res.con.acc} A, B and C show the error propagation
between the numerical calculations and the transformed results for
different grid resolutions. Although the error values are generally
small, it is visible that close to $\lambda=0$, where series
expansions are used, the local error values are higher. No specific
techniques were employed to reduce these errors, since this provides
an indication of the model's stability to local errors. The
convergence of errors are general and on higher grid resolutions the
difference in error values are less evident.

Figure \ref{sec:res.con.acc} D, displays the convergence behaviour
of the calculations. Comparing the error size of the $\rho$
calculations with the grid size, gives the order of convergence
between $1.7$ and $2.2$ i.e. around second order convergence. Since
no specific techniques were used to balance the error sizes in the
series expansion calculations with the CIVP calculations, the models
also shows good convergence sensitivity against local errors. This
effect will be investigated in detail in the future to determine how
the model will behave with initial data with realistic observational
error margins.

\clearpage

\section{Conclusion}
\label{sec:affine.conclusion}

The ideal observational approach proposed in \cite{ellis85} can
arguably be considered the most philosophically sound approach to
cosmology since it only studies the part of the Universe that is
causally connected to a cosmological observer. Practically, a very
precise map of the contents of the universe is required to develop
the full potential of this approach, which, even in the era of
precision cosmology, is not yet achievable. The more conventional
model based approach is therefore, and will probably for some time,
still be the most favourable approach for assessing our
understanding of the Universe. However, the model and observational
approaches do not have to be advocated in opposition since
restricted implementations of the observational approach can provide
insight into the question: Can our conventional understanding of the
Universe, which is based on a priori assumptions such as the
Copernican principle, be derived from less restrictive models making
only the most essential assumptions?

The affine CIVP model developed in this paper provides a numerical
implementation of the observational approach restricted to a dust
filled isotropic universe. The model is particularly well suited for
investigating the historic evolution of the observable universe as a
reversed CIVP evolved from observations on a local PNC. The line
element of the affine CIVP, is closely related to the Bondi-Sachs
metric, which makes it a direct process to adapt existing numerical
schemes to the new metric. As with the standard characteristic
formalism, second order convergence and accuracy was achieved when
verified against models with known solutions. Similar to the
simulations in \cite{vanderwalt10}, this provides a useful mechanism
to investigate the physical consistency of cosmological models given
identical observations on the observer PNC. With the standard
characteristic formalism, this could only be implemented in the
region prior to the AH, with the affine CIVP the region is extended
far enough to investigate the effect of the cosmological constant on
the history of the observable universe.

\acknowledgments
This work was supported by the National Research Foundation, South Africa.

\bibliographystyle{unsrt}
\bibliography{affine}

\begin{thebibliography}{10}

\bibitem{ellis85}
G.~F.~R. {Ellis}, S.~D. {Nel}, R.~{Maartens}, W.~R. {Stoeger}, and A.~P.
  {Whitman}.
\newblock {Ideal observational cosmology.}
\newblock {\em Phys. Reports}, 124:315--417, 1985.

\bibitem{uzan08}
{J.-P.} {Uzan}, C.~{Clarkson}, and G.~F.~R. {Ellis}.
\newblock {Time Drift of Cosmological Redshifts as a Test of the Copernican
  Principle}.
\newblock {\em Physical Review Letters}, 100(19):191303--+, May 2008.

\bibitem{hellaby09}
C.~{Hellaby} and A.~H.~A. {Alfedeel}.
\newblock {Solving the observer metric}.
\newblock {\em \prd}, 79(4):043501, February 2009.

\bibitem{temple38}
G.~{Temple}.
\newblock {New Systems of Normal Co-ordinates for Relativistic Optics}.
\newblock {\em Royal Society of London Proceedings Series A}, 168:122--148,
  October 1938.

\bibitem{araujo00}
M.~E. {Ara{\'u}jo} and W.~R. {Stoeger}.
\newblock {Exact spherically symmetric dust solution of the field equations in
  observational coordinates with cosmological data functions}.
\newblock {\em \prd}, 60(10):104020, November 1999.

\bibitem{araujo09}
M.~E. {Ara{\'u}jo} and W.~R. {Stoeger}.
\newblock {Obtaining the time evolution for spherically symmetric
  Lema{\^i}tre-Tolman-Bondi models given data on our past light cone}.
\newblock {\em \prd}, 80(12):123517, December 2009.

\bibitem{araujo10}
M.~E. {Ara{\'u}jo} and W.~R. {Stoeger}.
\newblock {Using time drift of cosmological redshifts to find the mass-energy
  density of the Universe}.
\newblock {\em \prd}, 82(12):123513--+, December 2010.

\bibitem{araujo11}
M.~E. {Ara{\'u}jo} and W.~R. {Stoeger}.
\newblock {Finding a spherically symmetric cosmology from observations in
  observational coordinates {--} advantages and challenges}.
\newblock {\em JCAP}, 7:29--+, July 2011.

\bibitem{musta98.1}
N.~{Mustapha}, C.~{Hellaby}, and G.~F.~R. {Ellis}.
\newblock {Large-scale inhomogeneity versus source evolution - Can we
  distinguish them observationally?}
\newblock {\em Mon. Not. R. Astron. Soc.}, 292:817--+, December 1997.

\bibitem{musta98.2}
N.~{Mustapha}, B.~A.~C.~C. {Bassett}, C.~{Hellaby}, and G.~F.~R. {Ellis}.
\newblock {The distortion of the area distance-redshift relation in
  inhomogeneous isotropic universes}.
\newblock {\em Classical and Quantum Gravity}, 15:2363--2379, August 1998.

\bibitem{lu07}
T.~{Hui-Ching Lu} and C.~{Hellaby}.
\newblock {Obtaining the spacetime metric from cosmological observations}.
\newblock {\em Classical and Quantum Gravity}, 24:4107--4131, August 2007.

\bibitem{mcclure08}
M.~L. {McClure} and C.~{Hellaby}.
\newblock {Determining the metric of the Cosmos: Stability, accuracy, and
  consistency}.
\newblock {\em \prd}, 78(4):044005, August 2008.

\bibitem{bish97}
N.~T. {Bishop}, R.~{G{\'o}mez}, L.~{Lehner}, M.~{Maharaj}, and J.~{Winicour}.
\newblock {High-powered gravitational news}.
\newblock {\em \prd}, 56:6298--6309, November 1997.

\bibitem{Winicour09}
J.~{Winicour}.
\newblock {Characteristic Evolution and Matching}.
\newblock {\em Living Reviews in Relativity}, 12:3, April 2009.

\bibitem{Reisswig:2009us}
C.~Reisswig, N.~T. Bishop, D.~Pollney, and B.~Szilagyi.
\newblock {Unambiguous determination of gravitational waveforms from binary
  black hole mergers}.
\newblock {\em Phys. Rev. Lett.}, 103:221101--1--221101--4, 2009.

\bibitem{bishop96}
N.T. Bishop and P.~Haines.
\newblock Observational {C}osmology and {N}umerical {R}elativity.
\newblock {\em Quaest. Math.}, 19:259--274, 1996.

\bibitem{vanderwalt10}
P.~J. van~der Walt and N.~T. Bishop.
\newblock Observational cosmology using characteristic numerical relativity.
\newblock {\em Phys. Rev. D}, 82(8):084001, Oct 2010.

\bibitem{hellaby06}
C.~{Hellaby}.
\newblock {The mass of the cosmos}.
\newblock {\em Mon. Not. R. Astron. Soc.}, 370:239--244, July 2006.

\bibitem{bondi60}
H.~Bondi.
\newblock Gravitational waves in general relativity.
\newblock {\em Nature}, 186:535--535, 1960.

\bibitem{bondi62}
H.~{Bondi}, M.~G.~J. {van der Burg}, and A.~W.~K. {Metzner}.
\newblock {Gravitational Waves in General Relativity. VII. Waves from
  Axi-Symmetric Isolated Systems}.
\newblock {\em Royal Society of London Proceedings Series A}, 269:21--52,
  August 1962.

\bibitem{sachs62}
R.~K. {Sachs}.
\newblock {Gravitational Waves in General Relativity. VIII. Waves in
  Asymptotically Flat Space-Time}.
\newblock {\em Royal Society of London Proceedings Series A}, 270:103--126,
  October 1962.

\bibitem{ellis71}
G.F.R Ellis.
\newblock Relativistic {C}osmology.
\newblock In {\em General {R}elativity and {C}osmology, {P}roc. {I}nt. {S}chool
  of {P}hysics `{E}nrico {F}ermi' ({V}arenna), {C}ourse {XLVII}, {E}d. {R}.{K}.
  {S}achs}, pages 104--179. Academic Press, 1971.
\newblock Reprinted in Gen. Rel. Grav. 41, 581 (2009).

\bibitem{Pasquini10}
L.~{Pasquini}, S.~{Cristiani}, R.~{Garcia-Lopez}, M.~{Haehnelt}, and
  M.~{Mayor}.
\newblock {CODEX: An Ultra-stable High Resolution Spectrograph for the E-ELT}.
\newblock {\em The Messenger}, 140:20--21, June 2010.

\bibitem{bish99}
N.~T. {Bishop}, R.~{G{\'o}mez}, L.~{Lehner}, M.~{Maharaj}, and J.~{Winicour}.
\newblock {Incorporation of matter into characteristic numerical relativity}.
\newblock {\em \prd}, 60(2):024005, July 1999.

\bibitem{burden93}
R.~L. Burden and J.D. Faires.
\newblock {\em Numerical {A}nalysis}.
\newblock PWS Publishing Company Boston, fifth edition, 1993.

\bibitem{ellis98}
G.~F.~R. {Ellis} and H.~{van Elst}.
\newblock {Cosmological Models (Carg{\`e}se lectures 1998)}.
\newblock In {\em NATO ASIC Proc. 541: Theoretical and Observational
  Cosmology}, pages 1--116, 1999.

\bibitem{Ribeiro95}
M.~B. {Ribeiro}.
\newblock {Observations in the Einstein-De Sitter cosmology: Dust statistics
  and limits of apparent homogeneity}.
\newblock {\em \apj}, 441:477--487, March 1995.

\bibitem{lidsey09}
J.E. Lidsey.
\newblock {ASTM}108 {C}osmology/{MTH703U} {A}dvanced {C}osmology.
\newblock Lecturer's notes.
\newblock \\ \url{http://www.maths.qmul.ac.uk/~jel/ASTM108/#coursenotes}
  (cited: January 2010).

\bibitem{lemaitre33}
G.~Lema\^{i}tre.
\newblock l'{U}nivers en expansion.
\newblock {\em Annales de la Soci\'{e}t\'{e} Scientifique de Bruxelles},
  53:51--+, 1933.
\newblock For an English transaltion see: The Expanding Universe. \emph{Gen.
  Rel. Grav}, 29(5):641, 1997.

\bibitem{tolman34}
R.C. Tolman.
\newblock Effect of inhomogeneity on cosmological models.
\newblock {\em Proc. Nat. Acad.}, Sci. 20:169, 1934.

\bibitem{bondi47}
H.~{Bondi}.
\newblock {Spherically symmetrical models in general relativity}.
\newblock {\em Mon. Not. Roy. Astr. Soc.}, 107:410, 1947.

\bibitem{celerier00}
{M.-N.} {C{\'e}l{\'e}rier}.
\newblock {Do we really see a cosmological constant in the supernovae data?}
\newblock {\em A \& A}, 353:63--71, January 2000.

\bibitem{pas99}
J.-F. {Pascual-S{\'a}nchez}.
\newblock {Cosmic Acceleration:. Inhomogeneity Versus Vacuum Energy}.
\newblock {\em Modern Physics Letters A}, 14:1539--1544, 1999.

\bibitem{enqv08}
K.~{Enqvist}.
\newblock {Lemaitre Tolman Bondi model and accelerating expansion}.
\newblock {\em General Relativity and Gravitation}, 40:451--466, February 2008.

\bibitem{gar08}
J.~{Garcia-Bellido} and T.~{Haugb{\o}lle}.
\newblock {Confronting Lemaitre Tolman Bondi models with observational
  cosmology}.
\newblock {\em JCAP}, 4:3, April 2008.

\bibitem{celerier07}
{M.-N.} {C{\'e}l{\'e}rier}.
\newblock {The Accelerated Expansion of the Universe Challenged by an Effect of
  the Inhomogeneities. A Review}.
\newblock {\em arXiv:astro-ph/0702416}, February 2007.

\bibitem{krasinski06}
J.~Pleba\'{n}ski and A.~Krasi\'{n}ski.
\newblock {\em An {I}ntroduction to {G}eneral {R}elativity and {C}osmology}.
\newblock Cambridge University Press, first edition, 2006.

\bibitem{araujo09.1}
M.~E. {Ara{\'u}jo} and W.~R. {Stoeger}.
\newblock {The angular-diameter distance maximum and its redshift as
  constraints on {$\Lambda$} {$\ne$} 0 Friedmann-Lema{\^i}tre-Robertson-Walker
  models}.
\newblock {\em Mon. Not. Roy. Astr. Soc.}, 394:438--442, March 2009.

\end{thebibliography}

\end{document}